\documentclass[acmtog, authorversion=True, nonacm=True, screen=true]{acmart}
\setcopyright{none}
\acmPrice{none}
\acmDOI{none}
\pdfoutput=1 

\AtBeginDocument{%
  \providecommand\BibTeX{{%
    \normalfont B\kern-0.5em{\scshape i\kern-0.25em b}\kern-0.8em\TeX}}}

\newcommand{\sellname}[0]{D-Flat}
\newcommand{\sellnameShort}[0]{D$\flat$}

\newcommand{\lx}[0]{x'}
\newcommand{\ly}[0]{y'}

\newcommand{\optparamSet}[0]{\Pi}
\newcommand{\optparamCell}[0]{\pi}
\newcommand{\optparam}[0]{\pi_i}
\newcommand{\paramDegree}[0]{D}
\newcommand{\compparamSet}[0]{\Psi}

\newcommand{\cellGrid}[0]{\chi}
\newcommand{\setReal}[0]{\mathbb{R}}
\newcommand{\setComplex}[0]{\mathbb{C}}

\newcommand{\phase}[0]{\theta}

\newcommand{\trans}[0]{A}

\newcommand{\lightState}[0]{\Theta}
\newcommand{\opticalModulation}[0]{M }
\newcommand{\wave}[0]{u}
\newcommand{\fwave}[0]{U}
\newcommand{\lr}[0]{\rho'}
\newcommand{\sx}[0]{x}
\newcommand{\sy}[0]{y}
\newcommand{\sr}[0]{\rho}
\newcommand{\width}[0]{w}
\newcommand{\mlp}[0]{\text{MLP}}
\newcommand{\mlpParams}[0]{w}

\newcommand{\imag}[0]{j}
\newcommand{\wv}[0]{k}
\newcommand{\wl}[0]{\lambda}
\newcommand{\resp}[0]{h}
\newcommand{\fresp}[0]{H}
\newcommand{\ft}[1]{\mathcal{F}\left(#1\right)}
\newcommand{\ift}[1]{\mathcal{F}^{-1}\left(#1\right)}
\newcommand{\hankel}[1]{\mathcal{H}\left(#1\right)}
\newcommand{\ihankel}[1]{\mathcal{H}^{-1}\left(#1\right)}
\newcommand{\image}[0]{I}
\newcommand{\scene}[0]{\mathcal{S}}
\newcommand{\psf}[0]{f}
\newcommand{\texture}[0]{T}
\newcommand{\point}[0]{p}
\newcommand{\weight}[0]{\omega}
\newcommand{\hx}[0]{\tilde{x}}
\newcommand{\hy}[0]{\tilde{y}}
\newcommand{\mask}[0]{M}

\newcommand{\sceneInfo}[0]{Z}

\newcommand{\prop}[0]{P}
\newcommand{\render}[0]{\text{render}}
\newcommand{\expnumber}[2]{{#1}\mathrm{e}{#2}}

\usepackage{xcolor,colortbl}

\begin{document}

\title{D-Flat: A Differentiable Flat-Optics Framework for End-to-End Metasurface Visual Sensor Design}

\author{Dean S. Hazineh}
\affiliation{%
  \institution{Harvard University}
  \streetaddress{150 Western Ave.}
  \city{Allston}
  \country{USA}
  }
\email{dhazineh@g.harvard.edu}

\author{Soon Wei Daniel Lim}
\affiliation{%
  \institution{Harvard University}
  \streetaddress{29 Oxford St.}
  \city{Cambridge}
  \country{USA}
  }
  
\author{Zhujun Shi}
\affiliation{%
  \institution{Harvard University}
  \streetaddress{29 Oxford St.}
  \city{Cambridge}
  \country{USA}
  }

\author{Federico Capasso}
\affiliation{%
  \institution{Harvard University}
  \streetaddress{29 Oxford St.}
  \city{Cambridge}
  \country{USA}
  }
\email{capasso@seas.harvard.edu}

\author{Todd Zickler}
\affiliation{
  \institution{Harvard University}
  \streetaddress{150 Western Ave.}
  \city{Allston}
  \country{USA}}
\email{zickler@seas.harvard.edu}

\author{Qi Guo}
\affiliation{
  \institution{Purdue University}
  \city{West Lafayette}
  \country{USA}
}
\email{guo675@purdue.edu}


\begin{abstract}
  Optical metasurfaces are planar substrates with custom-designed, nanoscale features that selectively modulate incident light with respect to direction, wavelength, and polarization. When coupled with photodetectors and appropriate post-capture processing, they provide a means to create computational imagers and sensors that are exceptionally small and have distinctive capabilities. We introduce \sellname~(\sellnameShort), a framework in TensorFlow that renders physically-accurate images induced by metasurface optical systems. This framework is fully differentiable with respect to metasurface shape and post-capture computational parameters and allows simultaneous optimization with respect to almost any measure of sensor performance. \sellnameShort~enables simulation of millimeter to centimeter diameter metasurfaces on commodity computers, and it is modular in the sense of accommodating a variety of wave optics models for scattering at the metasurface and for propagation to photosensors. We validate \sellnameShort~against symbolic calculations and previous experimental measurements, and we provide simulations that demonstrate its ability to discover novel computational sensor designs for two applications: single-shot depth sensing and single-shot spatial frequency filtering.
\end{abstract}



\maketitle

\section{Introduction}
\label{sec:intro}
\begin{figure}[t!]
    \centering
    \includegraphics[width=1.0\columnwidth]{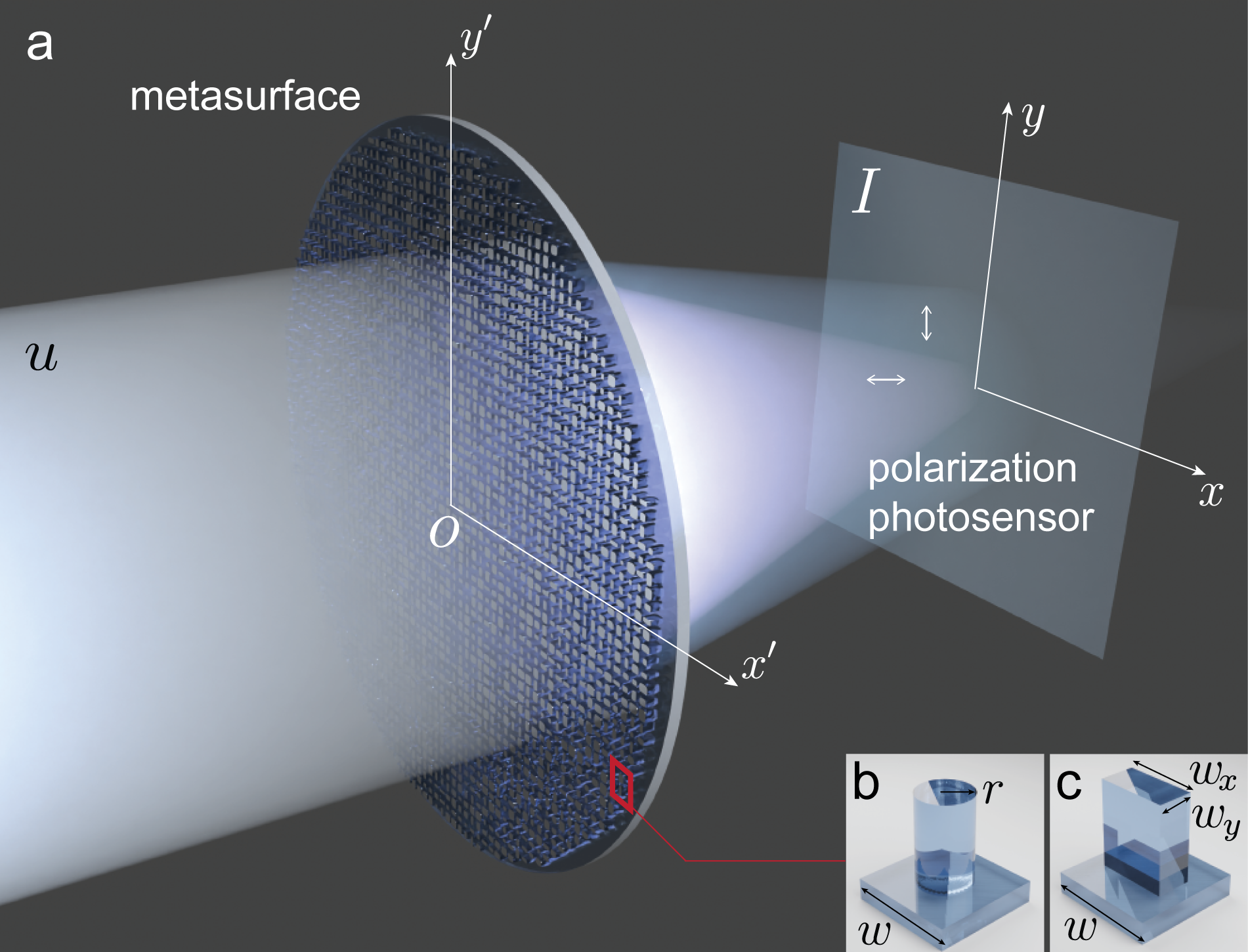}
    \caption{The metasurface plane is partitioned into sub-wavelength scale cells with independent shapes. The cells are approximated as being optically non-interacting. Common cell-shape dictionaries include (b) radius-parameterized nanocylinders and (c) width-parameterized nanofins. A typical width $\width$ of a cell is around $300$ nm for operation in visible light.
    Asymmetric cell-shapes like nanofins allow polarization control and can be designed across the plane to cooperatively induce two distinct images on orthogonal, linear polarization states ($\leftrightarrow$ and $\updownarrow$ in (a)), which may be captured simultaneously by a polarization-mosaicked photosensor.}
    \label{fig:metalens_depiction}
\end{figure}

Metasurfaces are a class of recently-matured, nanophotonic devices that consist of sub-wavelength scale structures patterned onto a planar transparent substrate. They have gained significant attention for their small size and their ability to enable \textit{custom multi-functionality}, with optical properties beyond those attainable by bulk material. Unlike refractive and diffractive optical elements, whose dispersion and birefringence are fixed upon the choice of material, the wavelength and polarization response of metasurfaces can be customized based on the local nanoscale shapes. Moreover, by cooperatively interleaving different nanoscale structures across a metasurface plane, one can simultaneously induce multiple behaviors that are distinct in their spatial location, spectral selectivity, and/or polarization composition. For example, Figure~\ref{fig:metalens_depiction}a depicts a metasurface focusing different polarization components of an incident wave to different focal lengths, creating two distinct images that may be captured simultaneously by a polarization-mosaicked photosensor.
By co-designing metasurface shapes and post-capture processing algorithms, researchers have recently demonstrated several small imagers~(e.g.,~\cite{tseng2021neural, huang2022full}) as well as a variety of ``single-shot'' computational sensors that can measure depth~\cite{tseng2021neural}, polarization~\cite{noah_polarization, lin_gruyter} or hyperspectral information~\cite{lin2021end} in exceptionally small form factors and without having to capture multiple exposures over time. 

In order to accelerate the pace of research in metasurface visual sensing, we present \sellname~(\sellnameShort), an open-source TensorFlow framework for the simultaneous optimization of metasurface shape and post-capture processing algorithms. \sellnameShort~is computationally efficient. It enables gradient-based, end-to-end optimization of one or more millimeter to centimeter diameter metasurfaces together with the post-capture processing parameters of a convolutional neural network (CNN) or any other differentiable computation. \sellnameShort~is designed to enable fast experimentation. Different computational-sensing models and architectures can be assembled quickly and easily using minimal code. We demonstrate in this paper the inverse-design of two distinct metasurface imaging systems optimized with different objectives. Lastly, \sellnameShort~is designed to be modular, providing developers with the ability to include future implementations of metasurface design and post-capture processing algorithms. 

To enable the inverse design of large-area metasurfaces, \sellnameShort~leverages a standard, cell-based approach \cite{generalizedSnellsLaw}: It partitions the metasurface into sub-wavelength cells of equal size, evaluates the optical response of the nanoscale shape in each cell independently, and then jointly propagates the field of per-cell responses to the photosensor or other imaging planes. This approach is an approximation that circumvents the computational intractability of solving directly for the electromagnetic field (with nanometer resolution) across the entire millimeter-scale metasurface\footnote{Recent domain decomposition methods achieved full-area field simulations but only for micron-diameter devices~\cite{lin2019overlapping}.}. Previous reports have shown that it is sufficiently accurate when neighboring cells contain similar types of structures (e.g. nanocylinders or nanofins as depicted in Figure \ref{fig:metalens_depiction}b-c)~\cite{pestourie_large_area_meta, first_metalens_paper}. In this way, the task of designing a metasurface becomes that of designing the nanostructures of each cell, treating each cell as a modular component, and spatially arranging the cells to achieve the desired modulation of the entire incident field.

\sellnameShort~currently supports two types of differentiable models to evaluate the optical response of each cell. These two approaches are complementary and are well suited for different design tasks or target metasurface sizes. First, it incorporates the auto-differentiable implementation of rigorous coupled-wave analysis (RCWA)\footnote{The theory of RCWA was originally introduced for nanophotonics by Lalanne and Silberstein \cite{lalanne2000}} published by Colburn and Majumdar~\cite{colburn2021inverse}. RCWA directly solves for the optical response of each cell under the locally periodic assumption by computing numerical solutions to Maxwell's equations. While its computational cost is high compared to the following approximate model, this method enables the design of complicated cell nanostructures. 

As an alternative to RCWA, we propose and demonstrate in this work an approximate, neural optical model which learns the mapping between the nanostructures of each cell and its optical response using a multi-layer perceptron (MLP). The neural optical model is as accurate as but more computationally efficient than the numerical solvers of Maxwell's equations, when the shape of the nanostructures in each cell is simple and can be parameterized using a low dimensional vector. We demonstrate that neural optical models can precisely localize the optical resonance caused by certain nanostructures while costing orders of magnitude fewer floating point operations (FLOPs) per cell than numerical solvers of Maxwell's equations. In this work, we also compare the usage of the MLP to alternate, differentiable models including elliptic radial basis function networks (ERBFs) and multi-variate polynomial regressions \cite{tseng2021neural}. 

Along with this paper, we present the framework as a validated, maintained open-source software accessible at \url{https://tinyurl.com/DFlatRepo}. We have included libraries of pre-trained neural optical models for common cell-shape families along with the optical response datasets to facilitate further investigations into implicit representations of optics. 

\section{Framework}
\label{sec:framework}
\begin{figure}[t!]
    \centering
    \includegraphics[width=\columnwidth]{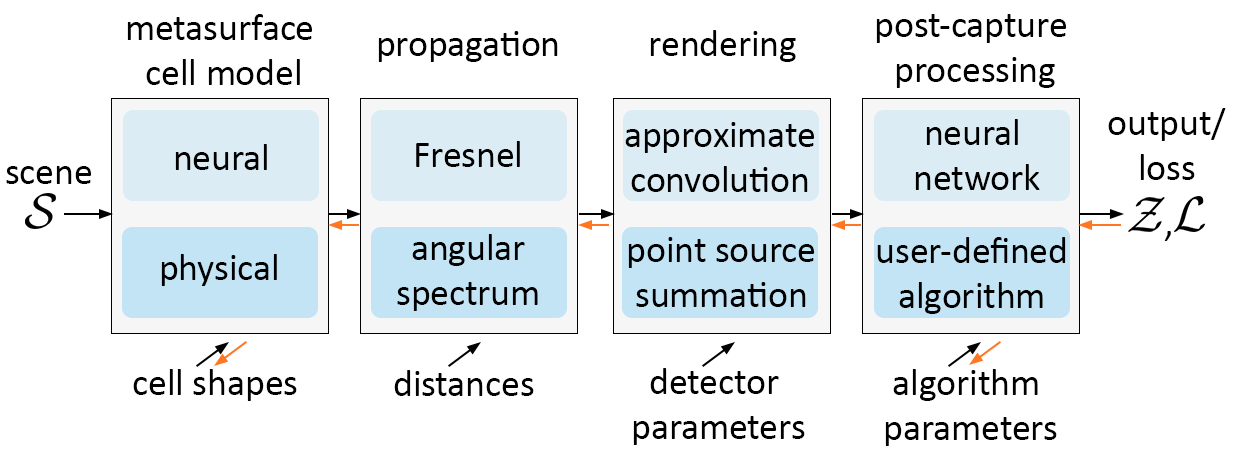}
    \caption{The \sellnameShort~ architecture consists of four basic stages and supports multiple implementations for each (see Section~\ref{sec:framework} for details). Each stage can be individually loaded as a modular, differentiable layer and layers can be chained together to simulate arbitrary imaging systems, including those incorporating multiple, cascaded metasurfaces. The aim of \sellnameShort~ is to provide a validated, comprehensive back-end to the metasurface, propagation, and rendering model so users may focus on experimenting with their own custom, post-capture processing and loss algorithms. Gradients can be back-propagated to train parameters in any layer.}
    \label{fig:pipeline}
\end{figure}

\sellnameShort~enables the user to differentiably render the image of a scene that is induced by one or more metasurfaces and measured by a photosensor. This is done through multiple, feed-forward computational layers as depicted in Figure~\ref{fig:pipeline}. First, it computes the complex modulation imparted by each metasurface cell onto a local portion of the incident wavefront (section \ref{ssec:optical_model}). It then jointly propagates the entire modulated field to the photosensor (section \ref{ssec:propagation_model}).  

Three-dimensional scenes $\scene$ are represented as sets of point-sources that cover the surfaces within the optic's field of view and reflect/emit light toward the system. The reflection/emission spectrum for each point-source may be explicitly defined with respect to wavelength and polarization. The image of the scene as captured on the photosensor can then be computed by evaluating and appropriately summing the optical system's response to each of the scene's sources. For 3D scenes comprised of textured, slanted planes at different depths, we incorporate an accelerated renderer that replaces the summation over point sources by approximate, piece-wise 2D convolutions of textured layers with the system's point-spread functions, as introduced by Guo et al. \cite{guo2019compact} (Section \ref{ssec:rendering_model}). 

Lastly, the rendered images may be passed to post-capture processing algorithms with trainable, computational parameters $\compparamSet$ to produce an output $Z$. Given examples of scenes $\scene$ and the desired sensor-outputs, $\{\scene_k; Z_{true, k}\}_{k=1,...,N}$,  \sellnameShort~enables the supervised co-optimization of the metasurface parameters $\optparamSet$ (e.g., geometrical dimensions of nanostructures) alongside the computational parameters $\compparamSet$ for different sensing tasks by solving the minimization problem,     
\begin{equation}
    \arg\min_{\optparamSet, \compparamSet} \sum_k \mathcal{L}\left( Z(\scene_k; \optparamSet, \compparamSet), Z_{true, k} \right).
\end{equation}
Here, $\mathcal{L}$ is a loss function to be minimized and measures the difference between the obtained output and the desired output. Examples of computational parameters $\compparamSet$ and optimization demonstrations with \sellnameShort~are provided in Section \ref{sec:experiment}.

\subsection{Metasurface Cell Models}
\label{ssec:optical_model}
At the heart of the proposed framework and one particular distinction which sets it apart from other diffractive simulators is a differentiable treatment of the metasurface optical response. We define a metasurface $\optparamSet$ as a collection of cells on a regular grid of points $\cellGrid$ at the metasurface plane. The nanostructures in each cell are then defined using a set of shape parameters $\optparamCell$:  
\begin{equation}
    \begin{split}
        \optparamSet &= \{\optparamCell(\lx, \ly) | (\lx, \ly) \in \cellGrid\}, \\
    \optparamCell(\lx, \ly) &= (\optparamCell_1, \optparamCell_2, ..., \optparamCell_i, ..., \optparamCell_D)\in \setReal^\paramDegree, 
    \label{eq:cellParameter}
    \end{split}
\end{equation}
where $\paramDegree$ is the dimensionality of the given nanostructure type. As an example, single nanocylinders placed at the center of each cell (Figure \ref{fig:metalens_depiction}b) are parameterized by the cylinder radius $r$ with $\paramDegree=1$, nanofins (Figure \ref{fig:metalens_depiction}c) by the fin widths $w_x$ and $w_y$ with $\paramDegree=2$, and the set of four ellipses (Figure \ref{fig:tps_2ndDeriv}b) by the set of four major and minor axes lengths with $\paramDegree=8$. In the limit of free-form nanostructures, $\paramDegree$ becomes particularly large as the parameters $\optparam$ may correspond to the binary inclusion of a dielectric at each point within the cell. 

An optical model $\opticalModulation:\setReal^\paramDegree \rightarrow \setComplex$ prescribes the mapping between the cell parameters $\optparamCell$ and its optical response in terms of a local transmittance $\trans$ and phase-delay $\phase$ imparted to an incident wavefront:
\begin{equation}
    \opticalModulation \left(\optparamCell(\lx,\ly), \lightState \right) = \trans(\lx, \ly) \exp\left(-\imag \phase(\lx, \ly) \right),
    \label{eq:opticalModulation}
\end{equation}
where $\lightState$ represents the state of the incident light defined by direction, wavelength $\lambda$, and polarization. The functionality of the assembled metasurface can then be defined by applying this complex modulation to the field incident at each cell
\begin{equation}
    \wave(\lx, \ly, 0^+, \lightState) =  \opticalModulation(\optparamCell(\lx, \ly), \lightState) \wave(\lx, \ly, 0^-,\lightState),
    \label{eq:opticalModel}
\end{equation}
where $0^-$ and $0^+$ represents the plane immediately before and after the optics, respectively.

Equation \ref{eq:opticalModel} is a statement of linear optics, while Equation \ref{eq:opticalModulation} is specific to metasurfaces. Unlike with conventional diffractive optical elements (DOEs), the cell size considered is sufficiently small relative to the operating wavelengths so that there is no energy in higher-order diffraction channels other than the zeroth order; as a result, only a single pair of transmission and phase values per state $\lightState$ needs to be modeled in the output \cite{generalizedSnellsLaw}. Moreover, as many structures have an optical response that varies weakly with incident angle, we may presume normal incidence. The optical model can be readily generalized to account for scenarios where either assumption is not valid, such as in non-local metasurfaces with a tilt-dependent response \cite{PhysRevLett.121.173004}. 

In order to optimize the metasurface under the constraint of modulation functions that are achievable by realistic nanostructures, the modulation function in Equation \ref{eq:opticalModulation} must be differentiable with respect to the shape parameters $\optparamCell$. Notably, there is no simple analytic relation for the optical mapping $\opticalModulation$ that may be derived from first principles which is valid for general nanostructures.

\subsubsection{Physical Optical Model}
\label{sssec:physical} 
Physical models determine the gradients of the mapping in Equation \ref{eq:opticalModulation} by solving Maxwell's equations directly, for a given cell. While there exists many methods (and many open source packages) for solving the field equations, including finite-difference (FDFD, FDTD) and finite-element (FEM) methods \cite{computational_em}, rigorous coupled-wave analysis (RCWA) \cite{s4} is largely the standard choice in meta-optics design owing to its computational efficiency, particularly for small cell sizes and high aspect ratio nanostructures. This approach avoids iterative solvers and Krylov methods by formulating the scattering problem as an eigenequation.

Given the ability to solve the forward equation, gradients in the reverse direction are typically obtained by the adjoint method \cite{OwenMiller, SJohnson_Adjoint}. The adjoint method yields directly an analytical solution to the gradients of the field with respect to dielectric inclusions at each point in the cell. Notably, we instead desire the gradients with respect to the shape parameters $\optparamCell$. While this can be done using the adjoint method, it requires the user to manually implement additional, shape-dependent derivations each time a new nanostructure type is introduced. Alternatively, this burden on users of the framework can be side-stepped entirely by employing an auto-differentiable (AD) field solver. Automatic differentiation stores the mathematical operations performed in a calculation so that the analytical chain rule for differentiation can be efficiently performed to yield the exact numerical gradients.

In recent years, there have been numerous AD field packages published. We modify and port the Tensorflow RCWA implementation \cite{colburn2021inverse} into~\sellnameShort. Technical details are deferred to the original work. In physical models, each cell must be individually discretized into a Cartesian grid to be numerically evaluated. Consequently, the required memory scales non-linearly with the resolution of the grid, and the computational cost for evaluating the mapping is generally found to be orders of magnitude greater than that of the neural optical model. For this reason, coupling a physical model for the optical layer with complex algorithms, like deep neural networks, is challenging. Alternatively, the computational cost is largely invariant to the value of $\paramDegree$ and requires no pre-evaluated training data.

\subsubsection{Neural Optical Model}
\label{sssec:neural}
The neural optical model is more efficient in inference than physical models and is similarly accurate. It uses trained MLPs to approximate Equation~\ref{eq:opticalModulation}:
\begin{align}
    \opticalModulation(\optparamCell,\lightState) \approx \mlp(\optparamCell, \lightState).
\end{align}
The MLP takes the cell parameters and the light state as an input and outputs the predicted transmittance and phase delay imparted by the cell. Given supervised training data 
$ \{\optparamCell; \opticalModulation(\optparamCell, \lightState)\}_{k=1,...,N}$ pre-generated by physical field solvers, we minimize the squared loss function to train the neural optical model via stochastic gradient descent:
\begin{align}
    \arg\min_{\mlpParams} \sum_k \sum_\lightState \left(\opticalModulation(\optparamCell, \lightState) - \mlp(\optparamCell,\lightState; \mlpParams)\right)^2,
    \label{eq:loss}
\end{align}
where $\mlpParams$ are the parameters of the MLP. 

While \sellnameShort~includes a compatible physical model (a Fourier-based method) to generate the training data, it is beneficial in some cases to utilize an efficient finite-difference time-domain (FDTD) solver instead \cite{fdtd}. Time-domain methods allow one to evaluate the optical response for many incident wavelength states with a single simulation by Fourier transformation of the time-domain field behavior. In contrast, frequency space methods require an additional simulation for each wavelength probed, although they are more efficient. For this work, we utilize the commercial FDTD software by Ansys Lumerical Inc., to generate several finely-sampled broadband datasets for training. 

The neural optical model serves as a differentiable proxy to the physical model, with substantially lower computational and memory cost per query. As a consequence, this representation enables the optimization of large, 2D metasurfaces in conjunction with complex algorithms--the limitation being the need to pre-generate training data. It is important to recognize, however, that the training data need not densely or uniformly sample all instantiations of $\optparamCell$ due to the  generalization power of MLPs. For large $\paramDegree$, e.g., multiple fins placed in a single cell with each having several degrees of freedom, it is possible to consider an adaptive learning method where the neural optical model queries the physical model for the labelled data that will best improve the models accuracy.

\subsubsection{Shape Constraints}
\label{sssec:constraints}
For both the physical model and the neural optical model, one must impose constraints when optimizing the shape parameters $\optparamCell$. While obtaining gradients with respect to the shape parameters rather than dielectric inclusions inherently ensures that only generally fabricable shapes are designed, we still require bounds on these degrees of freedom to ensure meaningful dimensions (e.g. structure widths that are positive and smaller than the cell size or separation distances consistent to fabrication tolerances). For the parameterizations discussed in this work, these bounds take the form of simple inequality constraints on the minimum and maximum dimensions, $a_i \le \optparam \le b_i$. To enforce these, we utilize a parameter-transformation method alongside standard unconstrained optimization. Specifically, for each cell, we back-propagate gradients to a latent parameter $z_i\in\setReal$ which is differentiably related to the shape parameters by the analytic transformation,  
\begin{equation}
   \optparam = a_i + (b_i-a_i)\frac{\tanh(z_i)+1}{2}, \optparam \in \optparamCell
\end{equation}
For more complicated and general constraints, this technique may still be used but with an alternate construction of the transformation function, potentially a pre-trained generator network or a set of nested functions.    

\subsection{Propagation Models}
\label{ssec:propagation_model}
Given the complex field after the optics, e.g. after applying Equation \ref{eq:opticalModel}, the propagation of the field in free space is prescribed fully by the theory of Fourier optics. In this section, the theory is reviewed with focus on the particular implementations included in \sellnameShort. In summary, \sellnameShort~incorporates four different propagation models for scalar fields which provide trade-offs between generality and efficiency. In addition to the full scalar diffraction model, it includes efficient propagators for metasurfaces with radial symmetry (e.g., metalenses) and for cases where the distance between the metasurface and photosensor is relatively large (i.e., paraxial propagation).

As in Figure~\ref{fig:metalens_depiction}a, the wavefront after the metasurface propagates towards the photosensor placed a distance $z$ after the lens. The field at the new plane, $\wave(\sx,\sy,z)$, can be computed by evaluating the first Rayleigh-Sommerfield solution of diffraction \cite{GoodmanFourierOptics4thEd}:
\begin{equation}
    \wave(\sx,\sy,z) = \iint_{\lx,\ly} \wave(\lx,\ly,0^+) h(x-\lx, y-\ly, z) d\lx d\ly.
    \label{eq:full_diff}
\end{equation}
The full transfer function $h(\sx,\sy,z)$ is the impulse response function for free space propagation:
\begin{equation}
    \begin{split}
        \resp(\sx,\sy,z) &= \frac{1}{\wl}\left(\frac{1}{\wv r} - \imag\right)\left(\frac{z}{r}\right)\frac{\exp{(\imag \wv r)}}{r}, \\
    \text{where }& r = \sqrt{\sx^2 + \sy^2 + z^2},\text{     } k = \frac{2\pi}{\wl}.
    \end{split}
\end{equation}

If $r\gg\lambda$, as is often the case in computational imaging, a binomial approximation (the Fresnel approximation) can be introduced and the impulse response function can be simplified:
\begin{align}
    h(\sx,\sy,z) = \frac{\exp{(\imag \wv z)}}{\imag \wl z} \exp{\left(\frac{\imag  \wv}{2z}(\sx^2 + \sy^2)\right)}.
\end{align}
This approximation enables a more computationally efficient form for Equation \ref{eq:full_diff} when substituted:
\begin{gather}
    \begin{split}
        \wave(\sx,\sy,z)&= \frac{\exp{(\imag \wv z)}}{\imag \wl z} \exp{\left(\frac{\imag k}{2z}(\sx^2 + \sy^2)\right)}\ \times\\
        &\ft{\wave(\lx,\ly,0^+) \exp{\left(\frac{\imag \wv}{2z}(\lx^2 + \ly^2)\right)}}_{f_{x'} = \frac{\sx}{\wl z}, f_{y'}= \frac{\sy}{\wl z}}\label{eq:fres}
    \end{split}
\end{gather}
which is referred to as the \textit{Fresnel diffraction integral}. In this approach only, the output grid differs from the input grid unless zero-padding of the initial field is used. This padding is handled internally in \sellnameShort~ to return fields on a user-specified grid. Evaluation of the diffraction equation can then be done utilizing a single Fourier transform operation. 

Notably, another computationally efficient form of Equation \ref{eq:full_diff} can be given which provides an exact treatment for the impulse response function. In some cases, it is also more memory efficient than Equation \ref{eq:fres}, as the input and output grids are the same without need for padding. In the frequency domain, Equation~\ref{eq:full_diff} can be reformulated as:
\begin{align}
    \fwave(\wv_x, \wv_y, z) = \fwave(\wv_x, \wv_y, 0^+) \fresp(\wv_x, \wv_y, z), 
\end{align}
where $\fwave$ denotes the Fourier transform of $\wave$ and $\fresp$ is the Fourier transform of the full $\resp$ \cite{Sherman_Fourier}:
\begin{align}
    \fresp(\wv_x, \wv_y, z) = \exp{\left(\imag z \sqrt{k^2 - \wv_x^2 - \wv_y^2}\right)}.
\end{align}
The propagated field can then be numerically evaluated via:
\begin{align}
    \wave(\sx,\sy,z) = \ift{\ft{u(\sx,\sy,0^+)}H(\wv_x, \wv_y, z)},
    \label{eq:ang_mom}
\end{align}
where $\ift{\cdot}$ indicates the inverse Fourier transformation. Equation~\ref{eq:ang_mom} is referred to as the \textit{angular spectrum method (ASM)}. 

To achieve acceptable accuracy with either propagation method, careful attention must be paid to the spatial sampling that is used in the discrete Fourier transforms. In particular, the initial field after the metasurface $\wave(\sx,\sy,0^+)$  should be sufficiently upsampled if needed to satisfy the Whittaker-Shannon sampling theorem. The required sampling rate can be deduced directly from the Fourier bandwidth of $\wave$; however, a challenge exists in that there is no simple, analytic theory to determine this bandwidth for arbitrary fields. In general, one would iteratively increase the sampling rate and evaluate the intensity of the field relative to the aliasing criteria (in practice, researchers often use an arbitrary upsample factor based on inspection). This is not suitable for inverse design where the field changes substantially and the computational graph for rendering should ideally remain fixed. 

To address this, we introduce an alternative method. In all cases, we instead consider the Fourier-bandwidth of a quadratic phase exponential bounded by the same field aperture as $\wave$. This bandwidth condition emerges naturally when the metasurface imparts an appropriate modulation to focus an incident field. An analytic relation for this Fourier bandwidth can then be derived and is dependent on $\lambda$ and $z$. Consequently, when dealing with multiple wavelengths or sensor distances, the computational graph is automatically branched with an appropriate, distinct upsample factor applied in each case. After the branched calculations are done, \sellnameShort~ then re-interpolates each field back to the user-defined grid. As a note, we find that if this condition on sampling were neglected, the predicted fields computed for the metasurface shown in Figure \ref{fig:HeartSingularity} would be incorrect.

When the wavefront $\wave$ is radially symmetric, i.e. $\wave(\sx, \sy,z) \equiv \wave(\sr, z)$ with $\sr = \sqrt{x^2 + y^2}$, the Fourier transforms in Equations~\ref{eq:ang_mom} and \ref{eq:fres} 
may be replaced with a Hankel Transform $\hankel{\cdot}$ to instead yield the propagation equations:
\begin{align}
    \wave(\sr, z) &= \ihankel{\hankel{\wave(\sr,0^+)}\hankel{h(\sr, z)}},
    \label{eq:ang_mom_r} \\
    \begin{split}
    \wave(\sr, z) &= \frac{\exp{(\imag \wv z)}}{\imag \wl z} \exp{\left(\frac{\imag k}{2z}\sr^2\right)}\ \times \\
    & \hankel{\wave(\lr,0^+) \exp{\left(\frac{\imag\wv}{2z}\lr^2\right)}}_{f_{\sr'} = \frac{\sr}{\wl z}}\label{eq:fres_r}
    \end{split}
\end{align}
For this work, we have introduced an auto-differentiable, approximate quasi-discrete Hankel transform based on the derivation in \cite{QDHT}.

\sellnameShort~implements the four different computations (Equation \ref{eq:fres}, \ref{eq:ang_mom}, \ref{eq:ang_mom_r}, \ref{eq:fres_r}) as auto-differentiable layers. Similar to the optical models, the most suitable or efficient choice for the propagator layer depends on the computational imaging task.

\subsection{Accelerated Rendering Model}
\label{ssec:rendering_model}
\begin{figure}[t!]
    \centering
    \includegraphics[width=0.9\columnwidth]{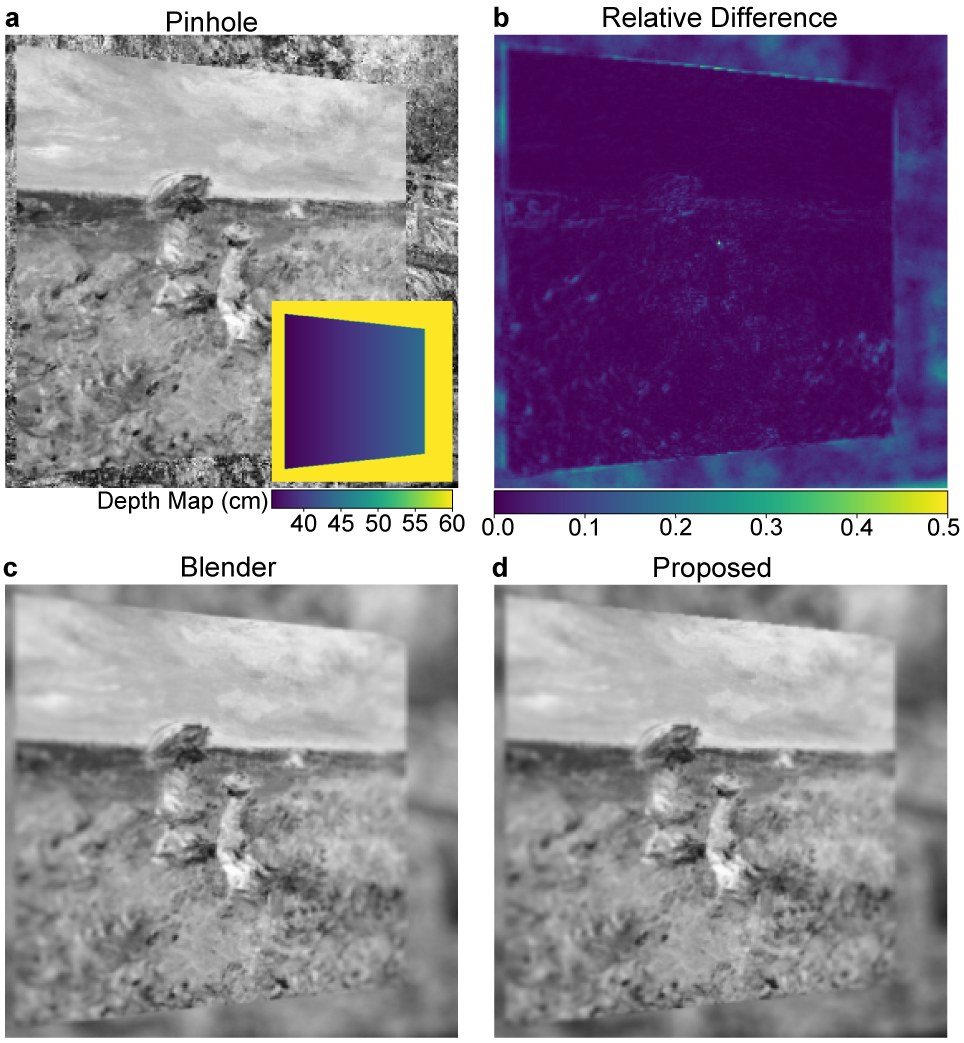}
    \caption{A comparison of the image produced by the accelerated renderer and by the commercial ray-tracing software, Blender. (a) The pinhole image of the scene. It consists of a slanted, textured foreground and fronto-parallel, textured background. The depth map is shown in the inset. (b) The relative difference between the rendered image by Blender (c) and by the proposed approach (d).}
    \label{fig:depth_blender}
\end{figure}

In this section, we briefly review the image formation model and the approximate rendering approach utilized in this paper. We treat the scene to be imaged as a collection of point light sources that are incoherent to each other (e.g., scattering under incoherent illumination). Without loss of generality, we may assume that the light is of a single wavelength and polarization, as a similar numerical process can be repeated to render images of the scene under different states.

Consider a scene $\scene$ where every point is completely visible to the optics.  Given that the transport of intensity is linear, the image $\image$ may be defined as the weighted summation of the intensity point spread functions $\psf$ (PSFs) produced by each point source on the surface of the scene, i.e. $\{ \point_i = (\sx_i, \sy_i, z_i)\in \scene\}$, such that,
\begin{align}
    \image(\sx, \sy) = \sum_{i} \texture_i \ \psf(\sx, \sy;\point_i),
    \label{eq:full_render}
\end{align}
where $\texture_i$ indicates the total energy of light emitted from the point source $\point_i$ that passes through the metasurface. Specifically, the PSF $\psf(\sx, \sy; \point_i)$ is the intensity distribution at the measurement plane, produced by a spherical wave which originated at the point $(\sx_i, \sy_i, z_i)$ and interacted with the optics. It can be numerically computed using the methods in Section \ref{ssec:optical_model} and \ref{ssec:propagation_model} via:
\begin{align}
    \psf(\sx, \sy; \point_i) = \left|\wave\left(\sx, \sy, z_s; \sx_i, \sy_i, z_i \right) \right|^2,
    \label{eq:PSF}
\end{align}
where $z_s$ is the distance from the metasurface to the photosensor.

For scenes that consists only of textured 3D planes (e.g. Figure~\ref{fig:depth_blender}a), we use an approximation to accelerate the rendering. We first assume that the memory effect holds such that the metasurface displays translational invariance for PSFs of the same depth. In other words, a point $\point_i=(\sx_i, \sy_i, z_i)$ produces a PSF equal to that from a source on-axis at the same depth but spatially shifted:
\begin{align}
    \psf(\sx, \sy; \point_i) = \psf(\sx - \hx_i, \sy- \hy_i; (0,0,z_i)),
\end{align}
where $\hx_i$ and $\hy_i$ are inhomogeneous coordinates of the point $\point_i$:
\begin{align}
    \hx_i = \sx_i z_s / z_i \text{ and } \hy_i = \sy_i z_s / z_i.
\end{align}
Furthermore, we assume that the PSFs vary slowly and smoothly with depth, such that we may calculate PSFs directly at only a few points ${p_j = (0, 0, z_j)}$ and approximate the others via linear interpolation:
\begin{align}
    \psf(\sx, \sy; \point_i) = \sum_{j} \weight^j_i\ \psf\left(\sx - \hx_i, \sy - \hy_i; p_j \right), \label{eq:interp}
\end{align}
where $\weight^j_i$ is the interpolation coefficient, dependent on the point $p_i$.

The image of a non-occluded scene $\scene$ can then be approximately rendered by combining Equation~\ref{eq:full_render}~and~\ref{eq:interp}:
\begin{align}
    \image(\sx, \sy) &= \sum_i \sum_j  \texture_i \ \weight^j_i\  \psf(\sx - \hx_i, \sy - \hy_i; \point_j),  \\
    &= \sum_j  \texture(\sx, \sy) \ \weight^j(\sx, \sy) * f(\sx,\sy;p_j), \label{eq:render}
\end{align}
where $*$ denotes a spatial convolution. The interpolation coefficient has been redefined via $\weight^j(\hx_i, \hy_i) = \weight^j_i$, and $\texture(\hx_i, \hy_i)$ corresponds to the pinhole image of the scene (an all in-focus, geometrically magnified image at the photosensor plane). While the linearly interpolated PSFs are an approximation to the true PSFs, the accelerated rendering closely resembles the synthesized result from ray tracing (see Figure~\ref{fig:depth_blender}). 

Lastly, we consider rendering scenes with occlusions, where certain points in the background are only partially visible to the optics. We render the depth boundaries using the layered-mask approximation in Guo et al. \cite{guo2019compact}. A scene $\scene$ is divided into layers $\{ \scene_l, l=1,2,3,...\}$, where points in each layer $\scene_l$ are not self-occluded. The framework then renders an image $\image_l(\sx, \sy)$ of each layer individually using Equation~\ref{eq:render} and the corresponding mask $M_l(\sx,\sy)$, an image of the segment with uniform texture, $T(x,y)=1$. The final image of the scene is synthesized by summing over the layer images and masks via:
\begin{align}
    \image(\sx, \sy) = \sum_l \image_l(\sx, \sy) \prod_{\text{$s$ occludes $l$}} (1 - \mask_s(\sx, \sy)). \label{eq:sum_layers}
\end{align}

\section{Validation}
\label{sec:validation}

\subsection{Optical Model Validation}
\label{ssec:neural_val}
For the demonstration of the neural optical model, we focus discussion here on the set of cells containing nanocylinders and nanofins (depicted in Figure \ref{fig:metalens_depiction}b,c). These two shapes have been two of the most commonly used nanostructures for designing metasurface-based imaging systems \cite{lin2021end,tseng2021neural, Chenreview}. Nanocylinders are polarization insensitive, while nanofins are polarization dependent due to their asymmetric shape. 

Importantly, for this validation, we desire a cell parameterization $\optparamCell$ for which it is feasible to densely sample the parameter space and evaluate the broadband, ground-truth optical response. Notably, although the nanofins have just two parameters ($\paramDegree=2$), assuming a fixed cell size and fin height, generating the dataset for a 350 nm wide cell utilizing FDTD as discussed in section \ref{sssec:neural} involved evaluating 2304 instantiations which took approximately 200 hours of parallel compute time on a 64-core server CPU. For cell parameterizations with larger values of $\paramDegree$, one would ideally query training data in a non-uniform, data-driven manner or utilize RCWA and sparsely sample incident wavelengths. 

\begin{figure*}[t!]
    \centering
    \includegraphics[width=1.0\textwidth]{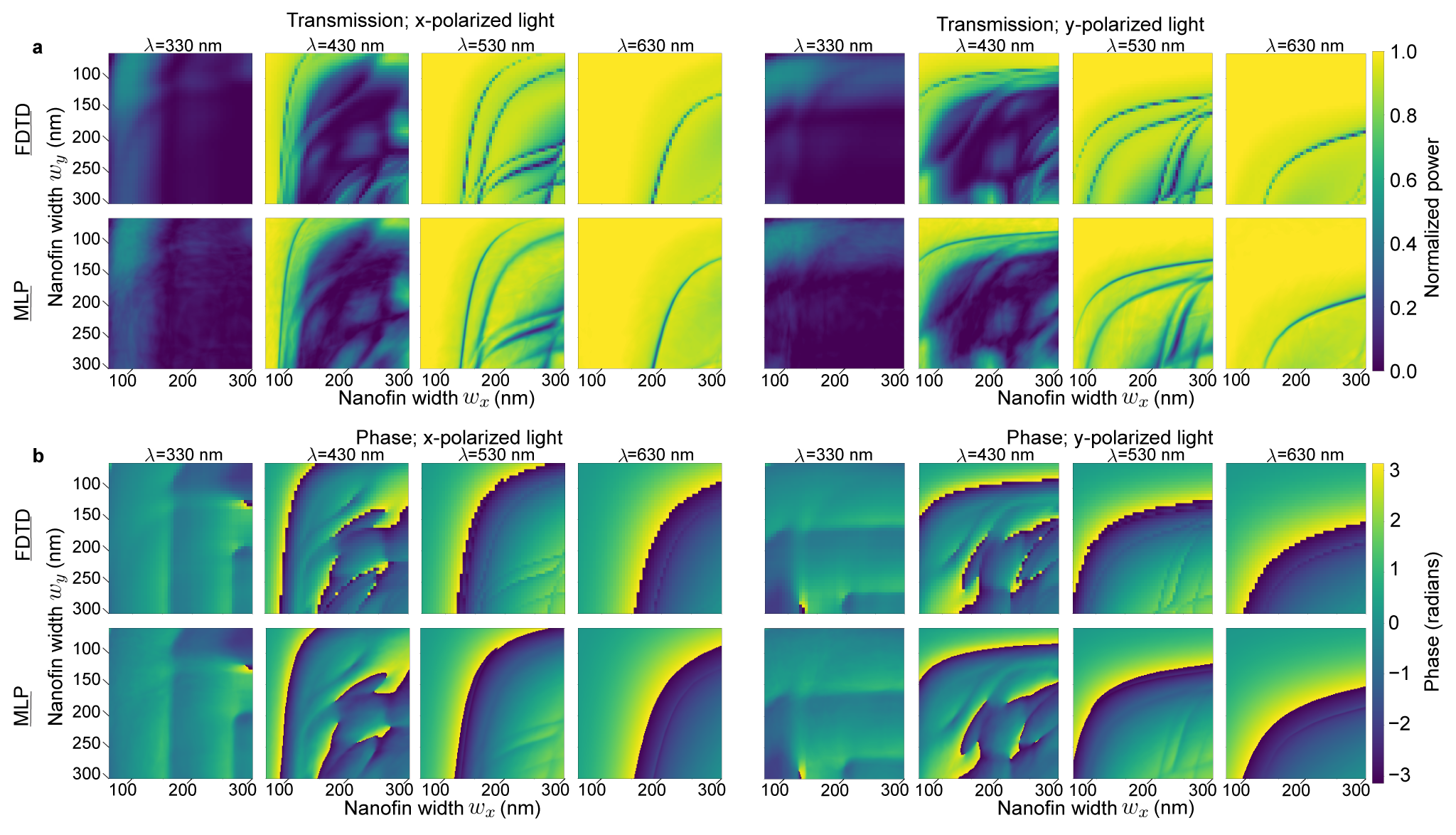}
    \caption{(Neural model: NO-D1024-fins) MLP-predicted transmission (a) and phase (b) imparted by a nanofin on a 350 nm cell for incident light linearly polarized along x and y. Inputs to the MLP are wavelength $\wl$ and the nanofin widths $w_x$ and $w_y$ (as depicted in Figure \ref{fig:metalens_depiction}c). The results are compared to the FDTD simulated dataset, which was generated by sweeping widths between 60 and 300 nm with a 5 nm step size and for wavelengths between 310 and 750 nm with a 1 nm step size. The MLP results are displayed for a grid of nanofin widths upsampled at 4x the resolution of the FDTD dataset.}
    \label{fig:Nanofin_mlp}
\end{figure*}

This FDTD-generated dataset for nanofins and the corresponding output of a trained neural optical model is displayed for several discrete wavelengths in Figure \ref{fig:Nanofin_mlp}. The model takes in fin widths $\width_x$, $\width_y$, and wavelength $\wl$ and outputs the predicted transmittance $\trans$ as well as the $\sin(\cdot)$ and $\cos(\cdot)$ of the phase delay $\phase$ for both x and y polarized light:    
\begin{align}
    \{\width_x, \width_y, \wl\} \rightarrow \{\trans_x, \sin{\phase_x}, \cos{\phase_x}, \trans_y, \sin{\phase_y}, \cos{\phase_y}\}
    \label{eq:mapping}
\end{align}
Dealing with the projection of the phase rather than the value directly is important to correctly handle the discontinuity in phase-wrapping. The displayed model used for this mapping has two hidden, dense layers with 1024 neurons in each (hence the name NO-D1024-fins in Table \ref{tab:optical_model_eval}). 

\begin{table}[t!]
    \centering
    \caption{Performance Per Cell Evaluation for Optical Models. ``NO" means neural optical model, ``D1024" indicates there are two hidden, dense layers with 1024 neurons in each, and ``fins/cylinder" represents the type of the cells. For the RCWA model, $(512^2, 121)$ represents a $512\times512$ grid for cells and 121 Fourier modes.}
    \label{tab:optical_model_eval}
    \small
    \begin{tabular}{|
    >{\columncolor[HTML]{EFEFEF}}l |
    >{\columncolor[HTML]{FFFFFF}}c |
    >{\columncolor[HTML]{FFFFFF}}c |
    >{\columncolor[HTML]{FFFFFF}}c |
    >{\columncolor[HTML]{FFFFFF}}c |}
    \hline
    Optical Model & \cellcolor[HTML]{EFEFEF}$\#$ Parameters & \cellcolor[HTML]{EFEFEF}FLOPs & \cellcolor[HTML]{EFEFEF} MAE Test Set\footnotemark \\\hline
    
    NO-D128-fins & 18 k & 37 k & 0.035\\ 
    
    NO-D256-fins & 68 k & 139 k & 0.025 \\ 
    
    NO-D512-fins & 276 k & 540 k & 0.021 \\ 
    
    NO-D1024-fins & 1.05 m & 2.13 m & 0.019 \\ \hline
    
    NO-D64-cylinder & 5 k & 10 k & 0.043 \\
    
    NO-D128-cylinder & 17 k & 36 k & 0.024 \\ 
    
    NO-D256-cylinder & 68 k & 138 k & 0.018 \\ \hline

    RCWA-($512^2,49$) & NA & 363.27 m & 0.062 \\ 
    
    RCWA-($512^2,81$) & NA & 1.620 b & 0.055 \\ 

    RCWA-($512^2,121$) & NA & 5.38 b & 0.051 \\ \hline

    \end{tabular}
\end{table}
\footnotetext{For RCWA, the MAE relative to FDTD is computed over cell transmission values only. The phase may have a global, constant offset without loss of accuracy. Due to computational limitations, the MAE is taken over a reduced sized set from the test cells.}

For this work, we tested different dense architectures to probe how the number of neurons in each layer affects the accuracy. While not displayed, we also explored changing the number of hidden layers; however, we found that two hidden layers provided sufficient accuracy while minimizing the number of trainable parameters. In Figures \ref{fig:supp_FinTrans_collage}-\ref{fig:supp_FinPhase_collage}, we show a similar display comparing the FDTD data against the nanofin MLP predictions as the number of neurons in each of the two hidden layers are reduced. The neural models for the nanocylinders are treated in the same way as the nanofins but with an input of $\{r,\wl\}$ and an output for just x-polarized light. A display of the nanocylinder MLP predictions are shown in Figure \ref{fig:supp_nanocylinderMLP}. All models utilize a leaky ReLU activation function and are trained on a desktop GPU with a MSE loss (Equation \ref{eq:loss}) and a standard Adam optimizer.

Empirically, we find that all neural optical models considered in this work are expressive enough to learn the general features of the optical response, for both nanocylinders ($D=1$) and nanofins ($D=2$). Moreover, some models are able to identify the cells that experience complex light-matter resonances. The resonances can be observed as the sharp dips in transmission in Figure \ref{fig:Nanofin_mlp}a. In practice, it is only important to identify the presence of these resonant cells (rather than to accurately characterize their optical response), since one will try to avoid their selection when designing metasurfaces to improve robustness to fabrication non-idealities. 

The accuracy of the neural optical model depends on the number of trainable weights in the MLP and the dimensionality $\paramDegree$ of the cells. While the method to query training data also has substantial influence, we consider only the case of uniform sampling of the cell's parameter space. To quantify the accuracy, we utilize as a metric the mean absolute error (MAE) of the complex modulation, taken over the testing set (cells which were not shown during training). The MAE for different models are listed in Table \ref{tab:optical_model_eval}, and we note that changes in this metric are found to be well correlated to the visual changes in accuracy that one can observe in Figures \ref{fig:supp_FinTrans_collage}-\ref{fig:supp_nanocylinderMLP}. As a benchmark, we also profile the auto-differentiable RCWA implementation included in \sellnameShort. For the RCWA calculations, cells are discretized into a $512 \text{ x }512$ Cartesian grid, the structure is assembled, and the electromagnetic fields are directly solved for. The accuracy of this calculation is set by the number of Fourier harmonics, and we consider 49, 81, and 121 modes (Table \ref{tab:optical_model_eval}). In practice, 121 modes are often used to obtain converged results. Evaluating the optical response of each cell by this method takes approximately $\expnumber{5}{9}$ floating point operations (FLOPs). 

In contrast, we note that the smallest nanofin model displayed (two hidden dense layers of 128 neurons) correctly predicts the general features of the optical response with disagreement localized around only some of the resonant cells. In this case, the model requires a factor of approximately $10^5$ times fewer FLOPs to evaluate the optical response of the cell as compared to the RCWA physical model. We find that the NO-D512-fins model (two hidden layers of 512 neurons) presents a nice balance between accuracy and computational cost. In summary, there is a trade-off between accuracy and computational advantage for the neural models and further optimizing this interplay for higher dimensional shapes is a topic of future investigation.

\subsubsection{Alternative Implicit Representations}
\label{sssec:implicit_rep}
As potential alternatives to the proposed neural optical model, we also consider and test the usage of other implicit representations to differentiably approximate the optical response of cells. Specifically, we evaluate the performance of a multivariate polynomial model, as introduced recently for metasurface end-to-end design by Tseng et al. (\cite{tseng2021neural}), and the performance of standard elliptic radial basis function networks (ERBFNs). To the best of our knowledge, the latter has not previously been applied in the context of this problem. 

The multivariate polynomial approach formulates the optical model in a linear, matrix form; the shape parameters and the incident wavelength are used as feature inputs and a coefficient matrix for each output is fitted to the FDTD data by the method of least squares. We consider different values for the maximum polynomial degree. The ERBFN is similar to the MLP approach in that both are a class of feed-forward neural networks with dense connections; however, the ERBFN instead utilizes a single hidden layer of neurons with an elliptic Gaussian activation function\footnote{Formally, they also differ in that the ERBFN computes a Euclidean distance between inputs and weights while the standard MLP architecture utilizes dot products}  \cite{KONONENKO2007275, DashBeheraDehuriCho}. The center coordinate and the standard deviations for the radial Gaussian activation function of each neuron is a learnable parameter that can be optimized via gradient descent, alongside the weights and bias of the dense connections to the output layer. The number of neurons (and consequently the number of radial basis functions used to represent the high-dimensional data) is a free-parameter. The predicted optical response generated by a representative polynomial model and ERBFN after fitting/training is displayed in Figure \ref{fig:supp_FinTrans_collage}b for the nanofin cells and in Figure \ref{fig:supp_nanocylinderMLP}b for the nanocylinders. In Figure \ref{fig:proxy_models}, we display the interquartile range of model errors and the corresponding computational cost of cell evaluation for ERBFNs of different size and for polynomial models of different order, with details provided in Table \ref{tab:supp_all_optical_models}. 

\begin{figure}[t!]
    \centering
    \includegraphics[width=1.0\columnwidth]{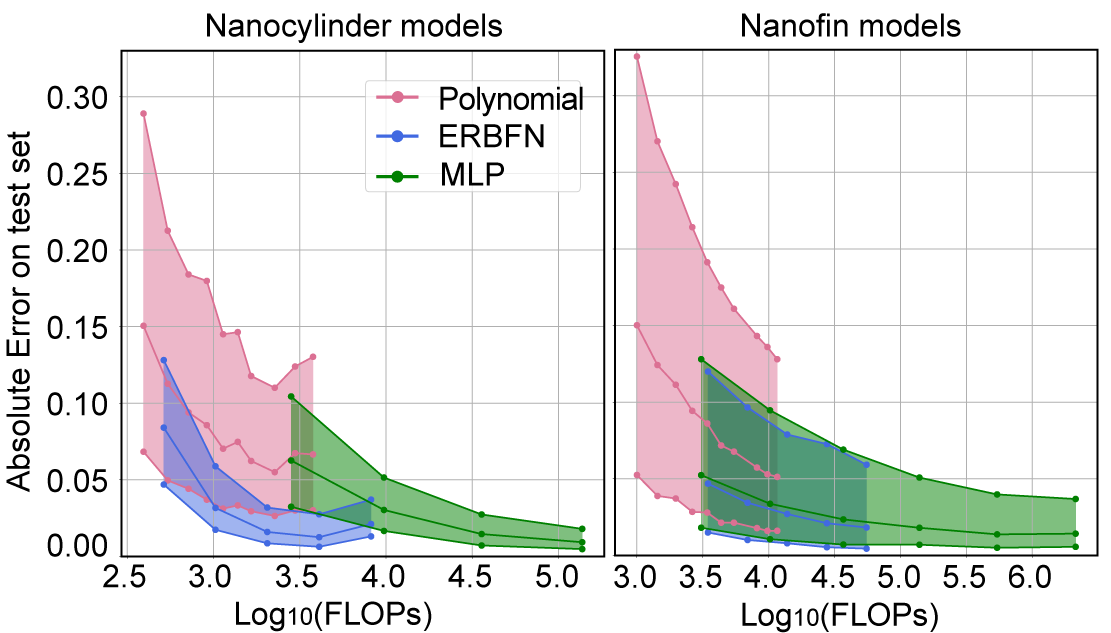}
    \caption{The median absolute error of cells' complex optical response is evaluated for each model. The interquartile range is displayed for the nanocylinder test set (left) and the nanofin test set (right). Different FLOPs for cell evaluations are obtained by changing the maximum order for the multivariate polynomial model and the number of neurons in the hidden layer(s) for the ERBFN and the MLP model. }
    \label{fig:proxy_models}
\end{figure}

For both types of cells considered in this paper, the polynomial model presents relatively large error in predicting the transmittance and phase. In the case of nanocylinders, the model begins to overfit with increasing polynomial order before converging to an accuracy comparable with RCWA. For nanofins, the large size of the training data and the higher dimensional input/output limited the maximum polynomial order that can be fitted by standard (non-iterative) regression methods to 15. In this case, the model is unable to localize any resonant cells and can only describe the general, low spatial frequency structures in the data.

In contrast, we find that the ERBFNs achieve comparable accuracy to the MLPs and present another reasonable approach for low dimensional cells shapes (i.e., $D=1,2$). For both nanocylinders and nanofins, the range of model errors when evaluating on the test set are similar, although the ERBFN has reduced computational cost for the nanocylinder cells relative to the MLP. On the other hand, we observe that the MLP can achieve better accuracy in identifying high spatial frequency features in the data which enables the MLP to predict resonant cells where the ERBFN fails (see ERBF-2048 in Figure \ref{fig:supp_FinTrans_collage}b as compared to Dense-512 in Figure \ref{fig:supp_FinTrans_collage}a as an example).  

While there are numerous variations to this study that could be done, e.g., tweaks to the ERBFN architecture, hyper-parameter tuning, or batched training for the polynomial regression, we highlight that the goal of the implicit representation here is not to absolutely minimize computation but to instead achieve a balance of computational cost, accuracy, and generality relative to auto-differentiable field solvers. While ERBFNs unsurprisingly perform well for these two shape families, the accuracy of this neural architecture requires having a number of hidden nodes roughly similar to the permutations of the input space, e.g., the curse of dimensionality \cite{Alpaydin14}. The neural optical models, however, perform well for both cases examined and are known to generalize better as the dimensionality increases for more complicated cell structures. Moreover, recent work on Fourier feature mapping by Tancik et al. \cite{tancik2020fourfeat} further suggests a route for improvements when using the more general, deep MLP architecture.

\subsection{Propagation Model Validation}
\label{ssec:prop_val}
To validate the differentiable propagators included in \sellnameShort, we first demonstrate its ability to reproduce in simulation the experimental fields measured by Lim et al. in \cite{SWDLim_Singularity_NatCommun_2021}. There, a metasurface was designed that produces a structured, 2D phase singularity at the sensor plane in the shape of a heart. A phase singularity is a point (or set of points) in a complex scalar field where the phase has a discontinuity making it undefined and these points are characterized by zero intensity. They are more commonly found in the unstructured form of optical speckle \cite{SpeckleSPIE}. Phase singularities make an ideal test-case for 2D propagation studies due to their sensitivity to field perturbations.

A scanning electron microscope (SEM) image of the fabricated metasurface is shown in Figure \ref{fig:HeartSingularity}a. In the leftmost panel (b) of the same figure, the experimentally measured intensity along with the reconstructed phase is displayed for the scenario where the metasurface is illuminated by a uniform plane wave and the photosensor is placed a fixed distance after the metasurface. Given knowledge of the metasurface shapes placed in each cell, we may define the field immediately after the metasurface according to Equation \ref{eq:opticalModel}. We then propagate this field and simulate a similar measurement at the sensor plane utilizing the implemented methods discussed in section \ref{ssec:propagation_model}. Here, we use both the 2D ASM method (Equation \ref{eq:ang_mom}) and the 2D Fresnel diffraction integral (Equation \ref{eq:fres}) and display the computed results in Figure \ref{fig:HeartSingularity}b. Additional comparison of the simulated and experimental measurements for different sensor distances are displayed in Figure \ref{fig:Supp_heart_hologram}. In all cases, we find good agreement between the experiment and the forward simulations.

\begin{figure}[t]
    \centering
    \includegraphics[width=0.48\textwidth]{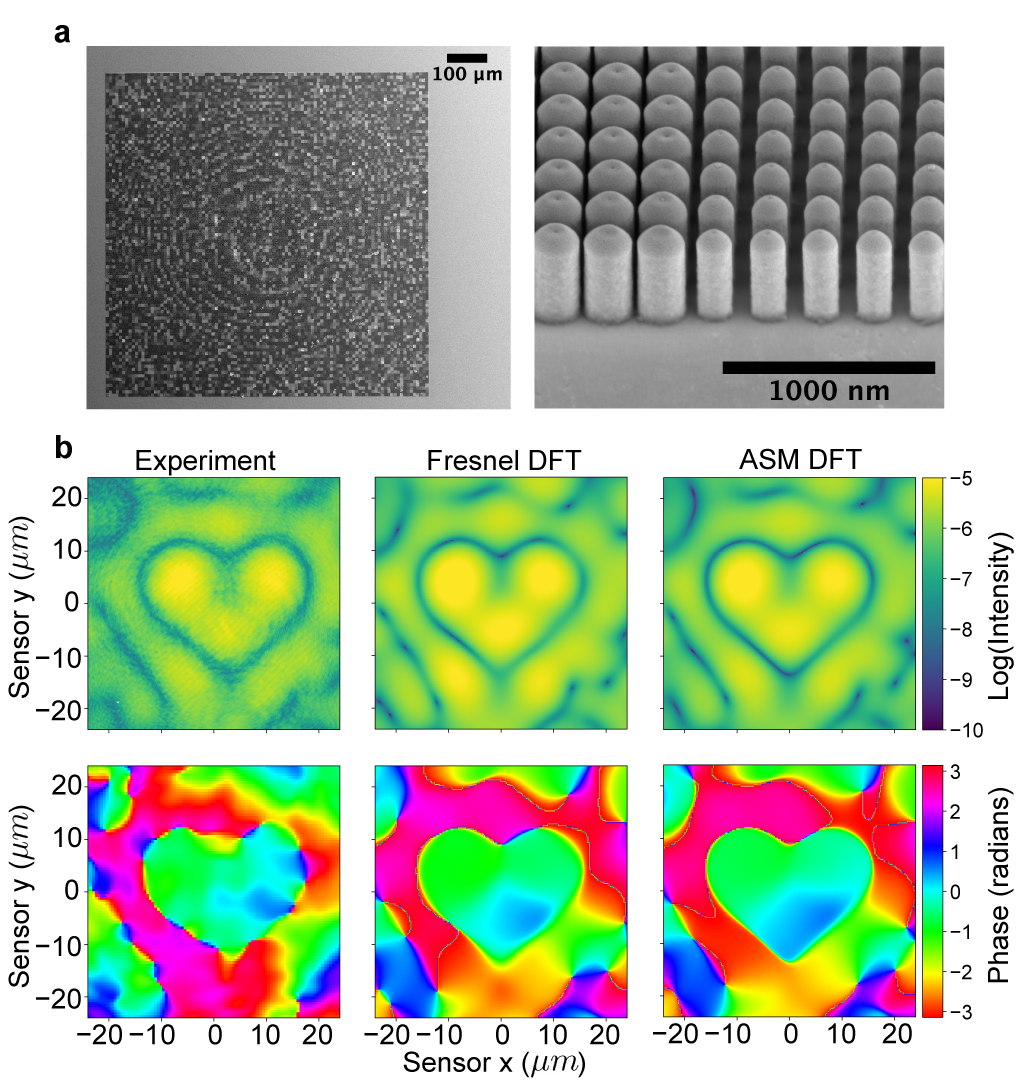}
    \caption{(a) SEM image of the fabricated metasurface from Lim et al. in \cite{SWDLim_Singularity_NatCommun_2021} built from 600 nm tall TiO$_2$ nanocylinders. (b) Left-most panel displays the experimentally measured intensity and the reconstructed phase at a sensor 9.8 mm after the metasurface. Center and right-most panels display the predicted intensity and phase computed by \sellnameShort~for the two implemented methods.}
    \label{fig:HeartSingularity}
\end{figure}

Alternatively, we also verify all four propagators released in this work by examining the computed intensity at the sensor plane for several metasurfaces designed to focus light. The required phase modulation needed to focus an incident plane wave to a diffraction-limited focal spot a distance $f$ after the metasurface is given by a hyperbolic phase function \cite{first_metalens_paper}:
\begin{align}
    \label{eq:focusing_phase}
    \phase(\lr) = \frac{2\pi}{\wl}\left(f-\sqrt{\lr^2 + f^2}\right). 
\end{align}
The resulting intensity distribution at the sensor should then match that prescribed by the Airy disk:
\begin{equation}
    \label{eq:airy_disk}
    I(\sr) = \left(\frac{2J_1(\chi)}{\chi}\right)^2 \text{; } \chi=\wv R \sin\left( \frac{\sr}{\sqrt{f^2 + \sr^2}}, \right)
\end{equation}
where $J_1$ is the Bessel function of the first kind of order one and $R$ is the radius of the circular aperture on the metasurface. The results of this analysis are shown in Figure \ref{fig:supp_metalens_focusing} with additional details in Appendix \ref{sec:supp_propagation_validation}. We find good agreement between the theoretical Airy profile and the computed intensity at the sensor for metalenses with different ratios of $R/f$ and using all four implemented propagators. When the ratio is larger than 0.3 (NA=0.29), the Fresnel approximation breaks down and as expected, only the ASM method is accurate.  

\begin{figure}
    \centering
    \includegraphics[width=0.50\textwidth]{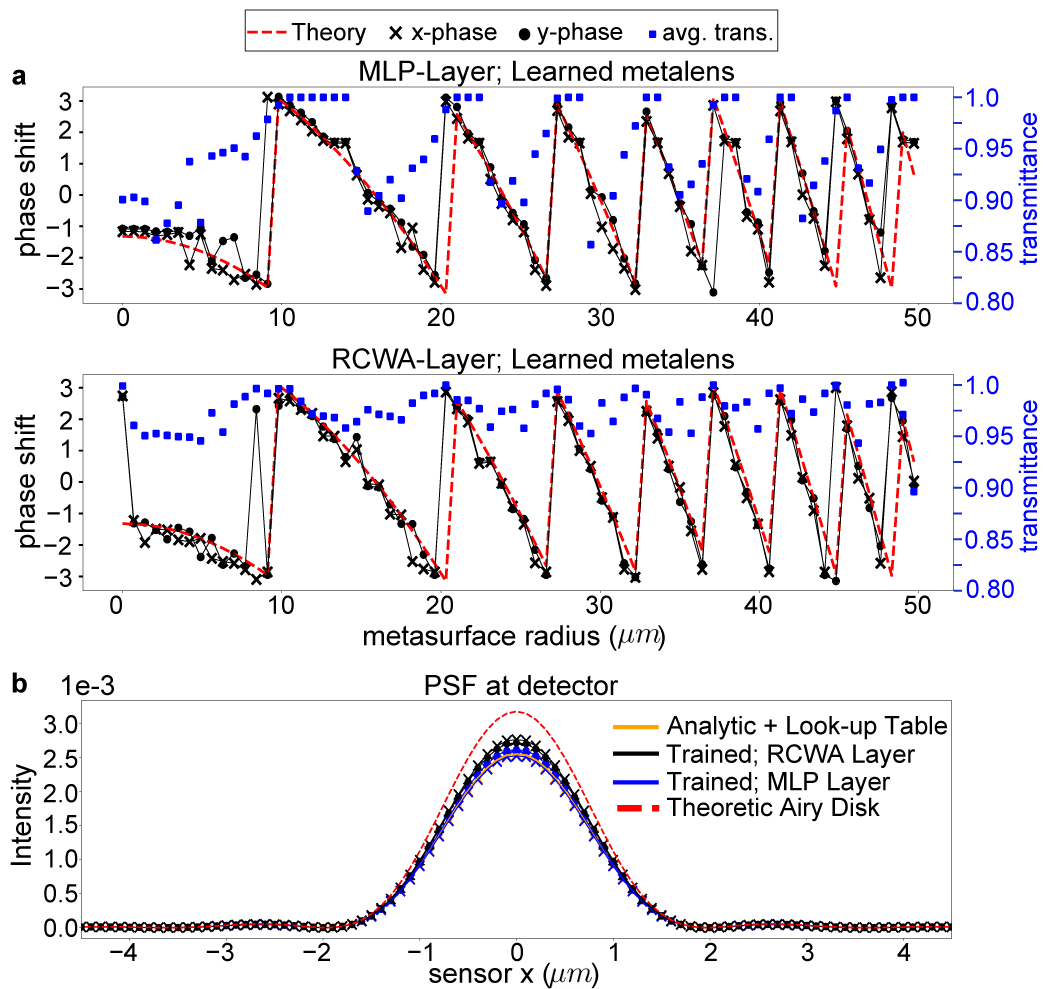}
    \caption{(a) The corresponding phase and transmittance imparted to x and y linearly polarized incident light by an optimized metasurface using the (top) neural optical model and (bottom) the RCWA physical model. (b) A radial slice of the intensity profile at the sensor plane for the two optimized metasurfaces along with the theoretic Airy disk profile. Intensity values correspond to a unity radiance incident plane wave.}
    \label{fig:optical_layer_valid}
\end{figure}

While the required phase profile for focusing with a metasurface is known a priori, we can instead combine the auto-differentiable optical layer and the propagator layer and probe by inverse design the metasurface $\optparamSet$ that the framework discovers for focusing light of a single wavelength. Due to memory limitations in utilizing the physical model as the optical layer for 2D calculations, we inverse design a radially-symmetric metasurface with the objective function:
\begin{align}
    \arg\max_{\optparamSet}  \sum_\lightState \prop\left( \wave(\lx,\ly, 0^-) \opticalModulation(\optparamSet, \lightState) \right) \delta(\sx=0,\sy=0).
\end{align}
$\prop(\cdot)$ denotes application of the propagator layer, here chosen to be the radial Fresnel method, $\sx$ and $\sy$ denotes coordinates on the sensor plane, and $\lightState$ corresponds to the incident, linear polarization states. In other words, we seek to maximize the intensity of the field measured at the center pixel of a photosensor. The results of this optimization are displayed in Figure \ref{fig:optical_layer_valid}. For the metasurface, we consider the placement of nanofin cells and utilize both the RCWA physical model and the neural optical model. Although not done here, $\lightState$ can be readily expanded to include incident wavelength for the inverse design of an achromatic focusing metasurface (all the models introduced in this work are broadband and the propagator layer calculations are efficiently batched over wavelength).

While nanofins generally produce polarization sensitive metasurfaces, the optimization in both cases learns to create a mostly polarization insensitive device by selecting only square nanostructures ($w_x$=$w_y$). The optimized phase profiles imparted by the designed metasurfaces are found to closely match the theoretic focusing profile in Equation \ref{eq:focusing_phase} (Figure \ref{fig:optical_layer_valid}a). Moreover, although only the intensity at the central pixel was maximized, the intensity distribution across the rest of the photosensor is found to converge to the shape of the analytic Airy disk (Figure \ref{fig:optical_layer_valid}b), demonstrating that the learning framework is physically consistent. Almost all energy that is incident on the metasurface is focused at the photosensor, although the peak intensity is slightly less than that predicted by the Airy disk since the transmittance of the placed cells are less than unity. The inverse-designed metasurface has a slightly better energy efficiency than that obtained by a metasurface forward-designed to implement the hyperbolic phase profile by dictionary look-up due to interpolation.

\section{Demonstration}
\label{sec:experiment}
In this section, we demonstrate via simulation the usage of \sellnameShort~to co-optimize a metasurface $\optparamSet$ in conjunction with a post-capture processing algorithm for two important visual sensing applications, incoherent opto-electronic image processing and snapshot depth sensing. Both applications exploit an optical setup similar to that depicted in  Figure~\ref{fig:metalens_depiction}a. 

A multifunctional metasurface produces two, unique PSFs carried on two orthogonal, linear polarization states ($0^\circ$ and $90^\circ$). A linear polarization-mosaicked photosensor, e.g., the SONY IMX264MZR photosensor, is then modeled to simultaneously capture the two images, $I_{0^\circ}$ and $I_{90^\circ}$, formed by the optics. For simplicity, we consider incident light of a single wavelength, $\lambda=532 \text{ nm}$; the simulations can be readily repeated for the broadband case. The width $\width$ of each metasurface cell is chosen to be $350$ nm, consistent to the trained neural optical models discussed in this work. We assume that the energy of $0^\circ$ and $90^\circ$ polarized light are the same in the incident field, which can be enforced in practice by adding a $45^\circ$ linear polarizer in front of the metasurface. The two rendered images, $I_{0^\circ}$ and $I_{90^\circ}$, are fed into different post-capture processing algorithms depending on the application. We use \sellnameShort~to differentiably model the optics and algorithm, and to jointly optimize the metasurface shape and the algorithmic parameters according to the objective function defined for each task.

\subsection{Single-Shot Incoherent Image Processing}
\label{ssec:image_differentiation}
In the optics community, there has been a long and rich history of designing optical systems for image processing \cite{LeeOpticalProcessing}. Numerous systems have been proposed and demonstrated for applying spatial frequency filters on a field, such that the captured image corresponds directly to a processed or spatially-differentiated rendering of the scene without any digital operations required\footnote{A simple example is the classic 4f imaging system where spatial frequency filtering is done optically by placing a mask at the Fourier plane.} \cite{PNASEdgeDetection, PhysRevLett.121.173004, ValentineImageDifferentiation}. Notably, these purely optical filtering methods all require that the scene is illuminated by coherent rather than incoherent light. This fundamental restriction can be understood by considering that an imaging system whose output is the derivative of the in-focus, incoherent image would require an intensity PSF with both positive and negative values--a condition which can not occur. Physically, the restriction is equivalent to the fact that the removal of signal cannot be achieved optically by destructive interference with incoherent addition of fields. 

To circumvent this limitation, we propose a new method based on extending the opto-electronic theory of two-pupil synthesis \cite{ChavelLowenthal, TwoPupilSynthesisRhodesLohman} to metasurfaces. The theory proposes that the removal of signal required for general spatial frequency filtering operations may be done digitally via the pixel-by-pixel subtraction of two images, captured on two co-designed optical systems. The benefit is a potentially substantial reduction in computational cost for image processing while the downside in practice stems from difficulties with photon noise. By leveraging the polarization-dependent optical response of the metasurface cells, the two required optical systems can be realized with a single optical component and more over, the two distinct images can be captured from the same perspective.

The net image which results from the digital subtraction of the two captured images, $I_{0^\circ}$ and $I_{90^\circ}$, may be given via: 
\begin{equation}
    \begin{split}
    \image_{0^\circ} &- \alpha \image_{90^\circ} = \texture * (\psf_{0^\circ}-\alpha \psf_{90^\circ})= \texture * \psf',
    \end{split}
    \label{eq:TPS}
\end{equation}
where $\texture$ is an all-in focus, magnified image of the scene, commonly referred to as a pinhole image. $\psf_{0^\circ}$ and $\psf_{90^\circ}$ are the two PSFs which form $\image_{0^\circ}$ and $\image_{90^\circ}$. All $\image$, $\texture$, and $\psf$ are functions of the detector spatial coordinate $(\sx, \sy)$, over which the convolution $*$ is applied. The coefficient $\alpha$ is introduced as a scalar constant that can be digitally applied and accounts for differences in total transmitted energy between the two PSFs. While each intensity PSF produced by the polarization multiplexed metasurface is individually positive, the digitally subtracted image corresponds to an effective PSF $f'(\sx, \sy)$ which may be positive and negative. In this sense, the digital output of the opto-electronic system can provide a spatially filtered or differentiated rendering of the scene for just two FLOPs per pixel. Moreover, given the checkerboard-mosaicked structure of the linear polarization filters at the photosensor, the scalar multiplication and the digital subtraction of the neighboring pixels signal can theoretically be done by analog circuitry enabling incoherent image differentiation without any digital, post-processing operations for the first time. 


\begin{figure}[t!]
    \centering
    \includegraphics[width=0.5\textwidth]{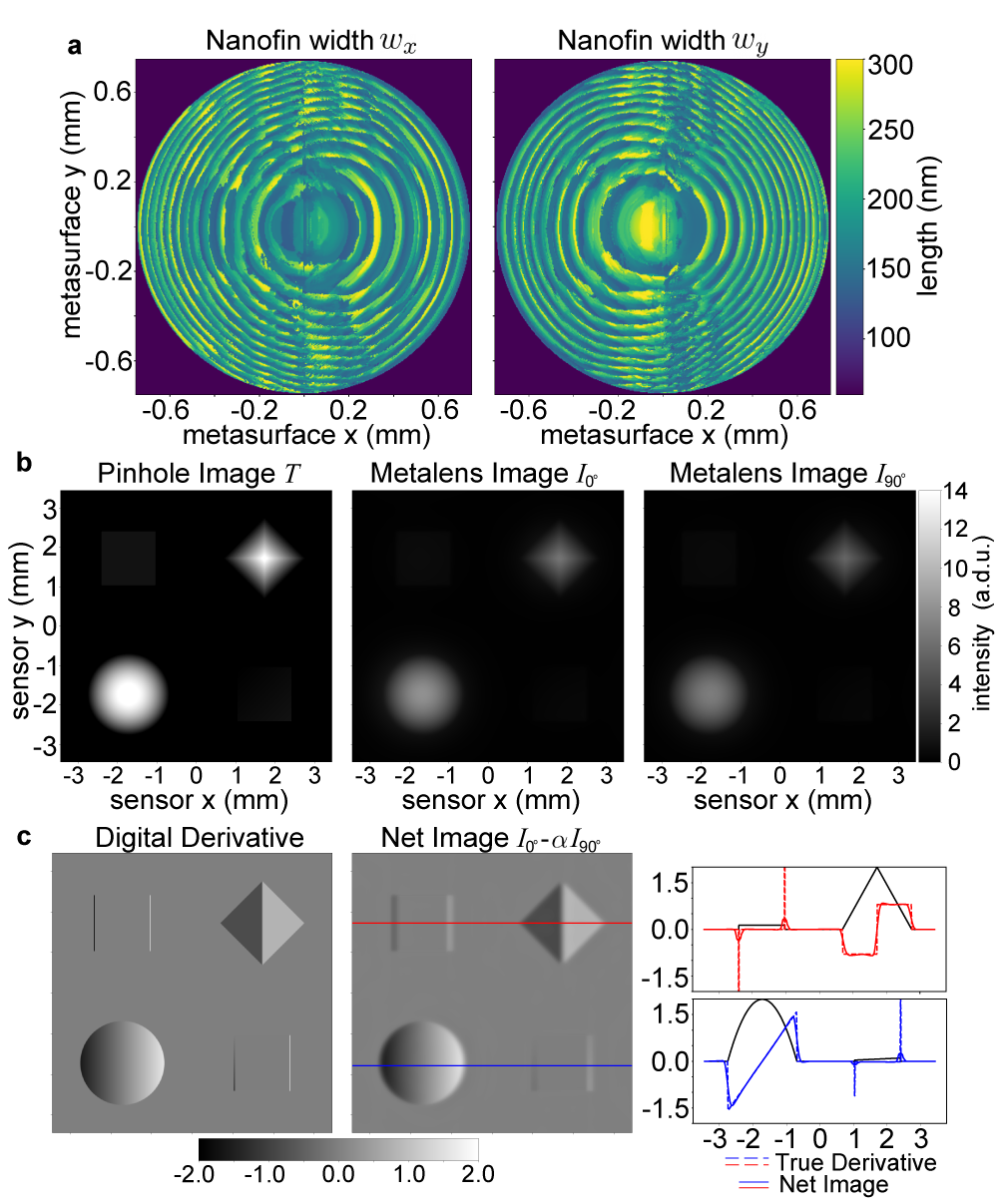}
    \caption{A neural optical model, pre-trained for nanofin cells, is coupled with the propagator and renderer to optimize a metasurface which enables capture of the first-derivative of a scene. (a) The optimized values for the nanofin widths $w_x$ and $w_y$ for each cell on the metasurface. (b) The pinhole image is displayed alongside the simulated measured images carried on the two, orthogonal polarization channels. (c) Digital subtraction of the two captured images with a learned scalar coefficient $\alpha$ approximates the true, first-derivative of the scene.}
    \label{fig:tps_firstDeriv}
\end{figure}

While $f'$ may be correlated directly to convolutional kernels that one may apply digitally, identifying the optics which realizes a given $f'$ is an inverse-problem that has no simple, analytic solution  \cite{PupilFunctionDesign,OTFSynthesis,JMait}. Instead, we use \sellnameShort~to learn an optimal metasurface $\optparamSet$ and scalar $\alpha$ for different filters and to design the functionality to be depth-invariant over a reasonable working range. To optimize the system, we require as an input the labeled data pair $(\scene_k, \sceneInfo_k)$, where $\scene_k$ is a series of fronto-planar scenes at different depths and $\sceneInfo_k$ is the corresponding, filtered image. The optimization is then done for a single wavelength by minimizing the L2 loss, starting from the initial condition of a uniform metasurface:
\begin{equation}
    \begin{split}
        &\arg\min_{\optparamSet, \alpha} \sum_k \left\|\left( I_1(\scene_k) - \alpha I_2(\scene_k) \right) - \sceneInfo_{k}\right\|^2 \\
        &I_i(\scene_k) = \render\left(\scene_k,  \opticalModulation(\optparamSet, \lightState), \prop \right), \quad \lightState \in \{0^\circ, 90^\circ\}.
    \end{split}
    \label{eq:TPS_argmin}
\end{equation}

We utilize this approach to design a metasurface-based opto-electronic system which produces in the net image a first-derivative rendering of the scene. For this example, we utilize a nanofin neural optical model and the 2D Fresnel integral propagator, and design the 1.5 mm diameter metasurface shown in Figure \ref{fig:tps_firstDeriv}a. The labeled training data consists of a scene corresponding to the displayed pinhole image and its first derivative along x, obtained by convolution with the 3x3 Sobel kernel (Figure \ref{fig:tps_firstDeriv}b,c). To enable depth invariant operation, we use the same labeled pair for three scene distances between 0.5 and 1.0 m in front of the metasurface. This optimization is carried out on a single, desktop Quadro RTX 4000 GPU using an Adam optimizer with a fixed learning rate of $\expnumber{1}{-3}$ and took approximately 40 minutes to train. We find that the net image produced by the co-designed metasurface matches well with the true first derivative (Figure \ref{fig:tps_firstDeriv}c). 

\begin{figure}[t!]
    \centering
    \includegraphics[width=1.0\columnwidth]{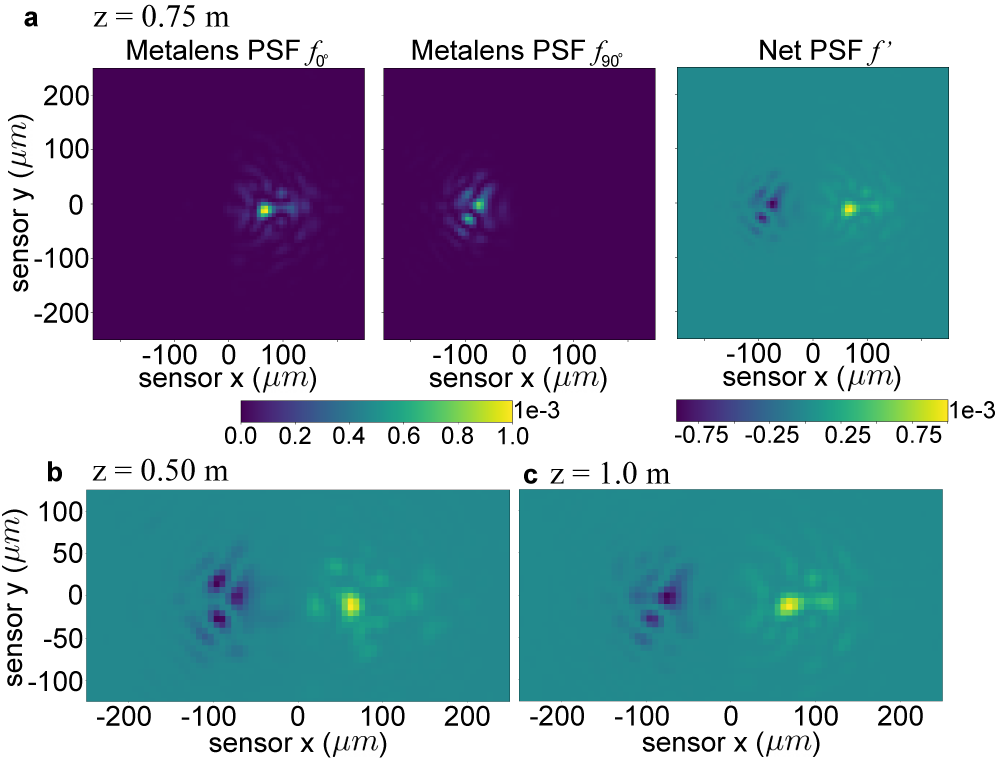}
    \caption{ (a) The PSFs measured on two, orthogonal linear polarization states, $0^\circ$ and $90^\circ$ at the photosensor, which is placed a fixed distance of 5 cm after the metasurface. The net PSF $\psf'$ is shown for three object to metasurface distances in the optimized range.}
    \label{fig:tps_firstDeriv_psf}
\end{figure}

The PSFs $\psf_{0^\circ}$ and $\psf_{90^\circ}$ for this optimized metasurface along with the net PSF $\psf'$ for the opto-electronic system is displayed for a scene depth of 0.75 m in Figure \ref{fig:tps_firstDeriv_psf}a. Notably, the net PSF learned by the optimization is antisymmetric with negative values in the left half and positive values on the right, matching the structure of the Sobel kernel that was used to generate the labeled data but was otherwise hidden to the optimizer. This suggests that the learned system can perform image differentiation to arbitrary scenes even when trained on a single example. Moreover, we find that the net PSF is largely depth invariant over the trained range (\ref{fig:tps_firstDeriv_psf}b,c), facilitating usage in real-world applications.   
 
The learned functionality can also be better understood by examining the optical transfer function (OTF) of the system, the frequency-space analogue of the PSF. A review of the OTF is given in Appendix \ref{sec:supp_tps_details} with the OTF for the system displayed in Figure \ref{fig:sup_tps_Firstderiv}. Here, we find that the transfer function of the optimized, opto-electronic system matches the definition of the first derivative (up to a proportionality constant) for a range of spatial frequencies about $f_x=0$. Disagreement occurs for high spatial frequency components and is unavoidable given the finite frequency bandwidth of the optics. This limitation introduces the error seen for the squares in Figure \ref{fig:tps_firstDeriv}c; however, the accuracy of the operation can be improved by designing a metasurface with a larger diameter. 

Extending from the previous discussion, we also show that the RCWA physical model may be used as the optical layer to solve a similar image processing problem but with a higher dimensional cell parameterization of $\paramDegree=8$. Using the same objective function of Equation \ref{eq:TPS_argmin} but re-targeted to a labeled training data pair corresponding to a second derivative, we optimize the radially symmetric metasurface shown in Figure \ref{fig:tps_2ndDeriv}a,b. The ground-truth, differentiated image is obtained by convolving the pinhole image with the 3x3, discrete Laplacian kernel. Rather than simple nanofins, we consider the placement of four ellipses in each cell, with each ellipse having a trainable major and minor axis length. The physical model then solves Maxwell's equations at each cell and for each training step and the auto-differentiable framework yields the gradients of the loss with respect to the shape parameters. A standard Adam optimizer with a fixed learning rate of $\expnumber{1}{-2}$ is used and the training takes approximately 40 minutes on the same desktop GPU. 

\begin{figure}[t!]
    \centering
    \includegraphics[width=0.5\textwidth]{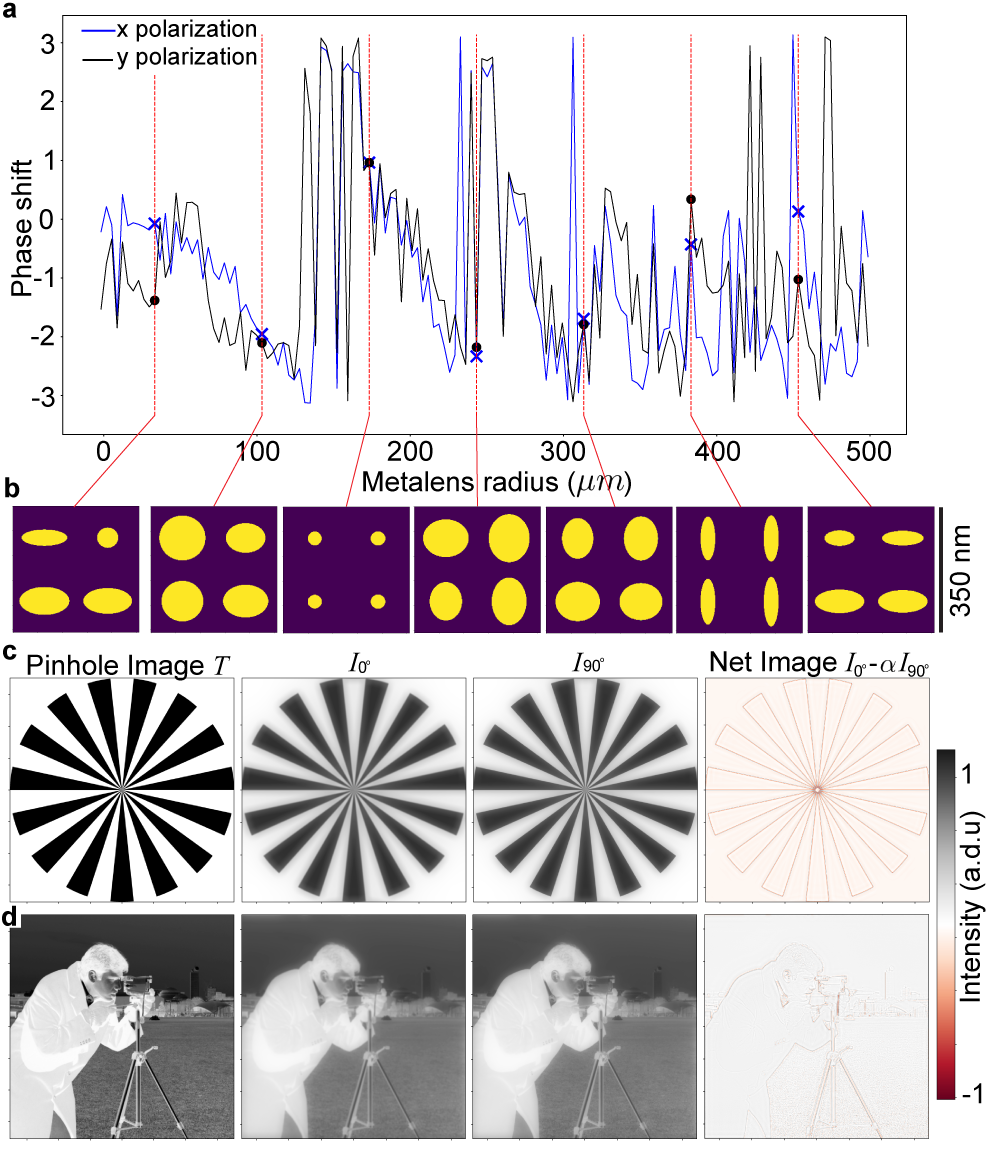}
    \caption{The RCWA physical model is coupled with the radial propagator and renderer to optimize a metasurface which enables capture of the second-derivative of a scene. (a) the corresponding phase modulation imparted by the optimized $\optparamSet$. A different phase delay is experienced for $0^\circ$ and $90^\circ$ linearly polarized light. (b) a top-down view of the optimized cells at select radial locations along the lens. (c) The pinhole image used for training is shown alongside simulations of the images produced by the metasurface and captured on the photosensor. The subtraction of the two images is done digitally post-measurement. (d) same as c, but for an unseen test image.}
    \label{fig:tps_2ndDeriv}
\end{figure}

In panel (c), the pinhole image used to train the metasurface (an image of a binary Siemens star) is displayed alongside the two photosensor images and the net image produced by the optimized opto-electronic system. To highlight the generality, panel (d) displays the simulated results for a test image. In both cases, we again find good agreement and observe that the metasurface-based imaging system learns to work in conjunction with the simple algorithm to produce a differentiated image suitable for edge-detection. Enlarged images are shown in Figure \ref{fig:sup_tps_2nd_deriv_zoom}a. In Figure \ref{fig:sup_tps_2nd_deriv_zoom}b, the optimized PSFs for the metasurface is displayed, and we observe that the net PSF $\psf'$ matches the shape of a Laplacian of Gaussian (LoG) kernel. For both optimization scenarios discussed here, we find that the framework is able to learn the underlying image filter by only looking at a single example. 

\subsection{Single-Shot Depth Sensing}
\label{ssec:depth_sensing}
\begin{figure}[t!]
    \centering
    \includegraphics[width=1.0\columnwidth]{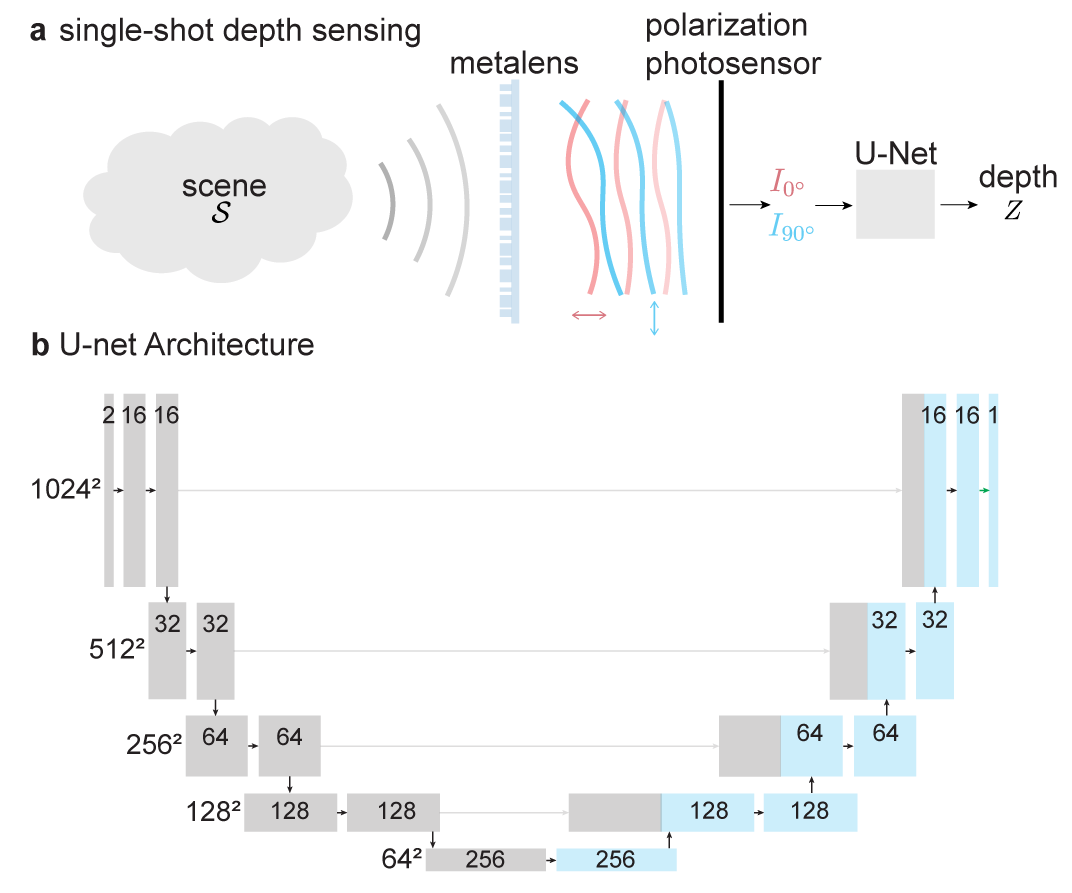}
    \caption{(a) Schematic depiction of the computational imaging architecture for single shot depth sensing. A metasurface produces two distinct, optical responses on two, linear orthogonal polarization states. The two images of the scene are formed and measured on the polarization mosaicked photosensor pixels. The distance between the metalens and the photosensor is set to 40 mm. The two images are then passed to a U-net which produces a depth map of the scene. (b) The U-net architecture used in this work. Rectangles denote 2D dense convolutional layers with 3x3 kernels, a ReLu activation function, and channel depth denoted by the overlaid number. Down arrows between blocks utilize a 2x2 maxpooling while the up arrows denote 2x2 upsampling.}
    \label{fig:depth_pipeline}
\end{figure}

\begin{figure}[t!]
    \centering
    \includegraphics[width=\columnwidth]{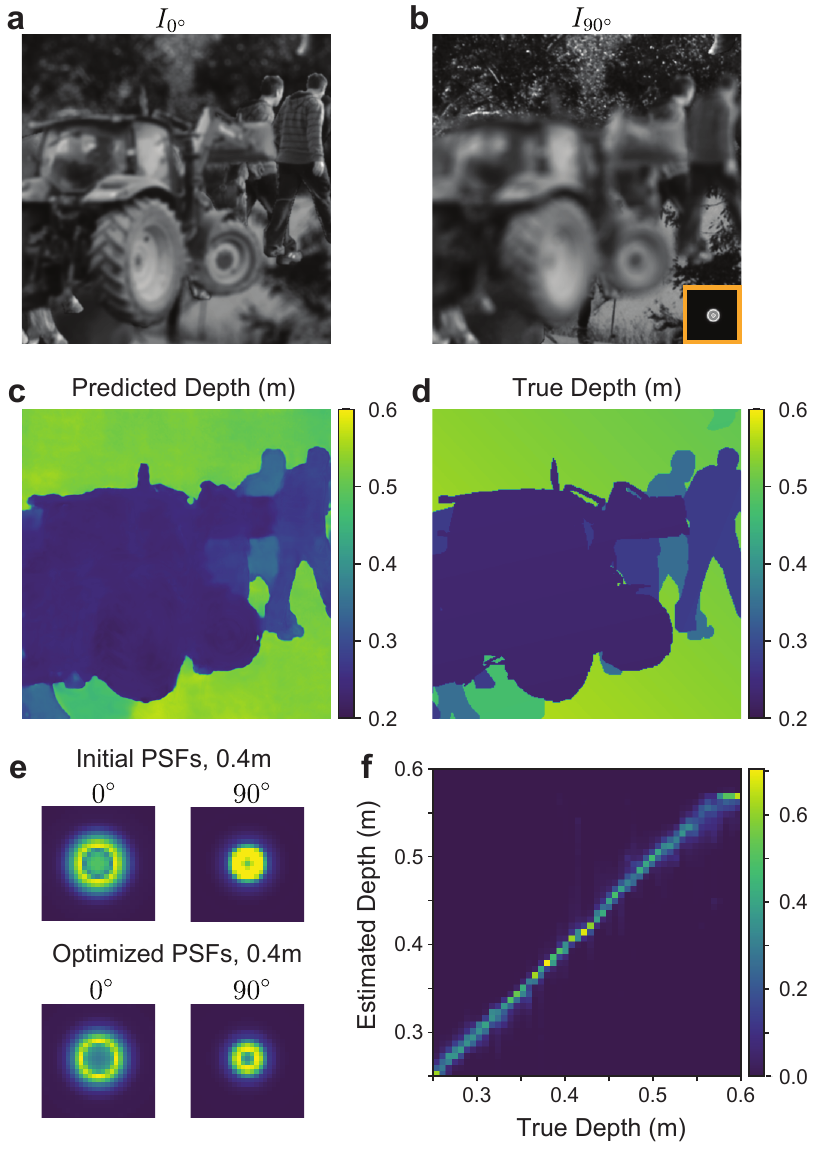}
    \caption{Single shot depth sensor utilizing the co-design of a metasurface and a U-Net. The simulated system simultaneously captures two differently blurred images (a,b) of a scene using the setup in Figure~\ref{fig:depth_pipeline}. It then processes the two images using a U-Net, which outputs the predicted depth map (c) of the scene. The receptive field of the U-Net (orange box in (b)) is relatively small compared the scale of image features, thus the network has to infer depth from local features such as image defocus, instead of from global, semantic features. During training, the metasurface cells are jointly optimized with the U-Net parameters. The PSFs of the system before and after the optimization at a sample depth is shown in (e). A 2D histogram of the predicted depth vs true depth on the 50 image testing set is displayed in (f).}
    \label{fig:depth-sensing}
\end{figure}

\begin{figure}[t!]
    \centering
    \includegraphics[width=1.0\columnwidth]{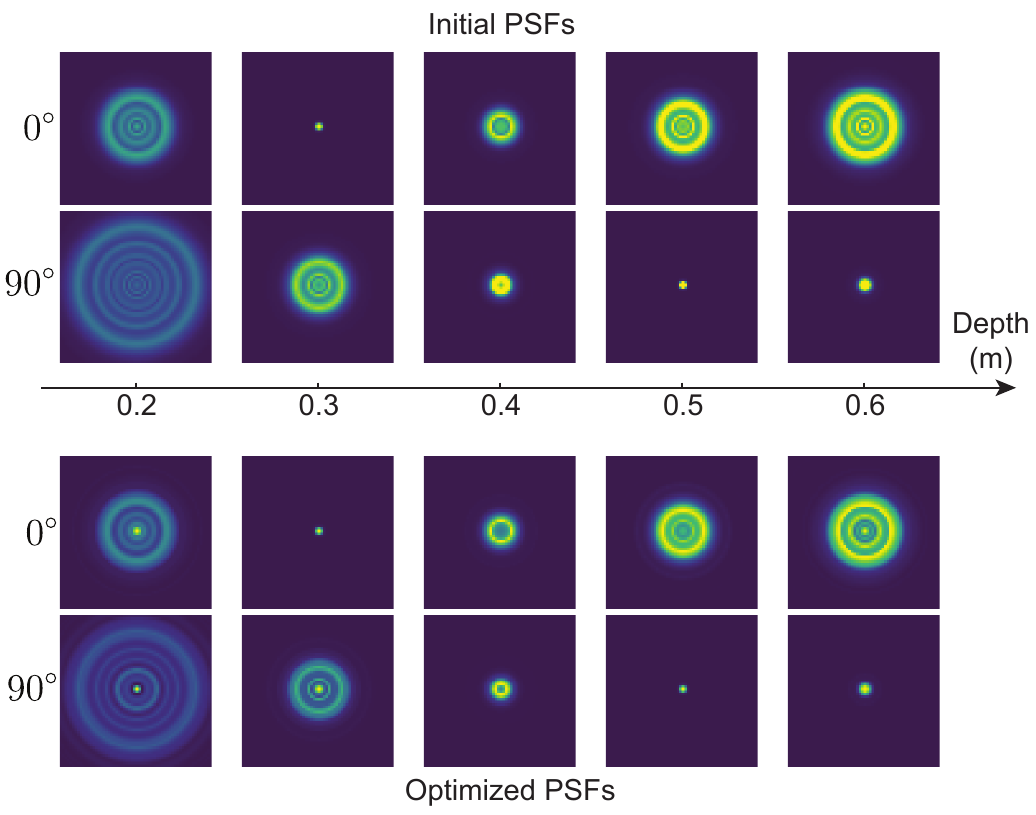}
    \caption{The simulated, initial (top) and trained (bottom) PSFs for the metasurface used in the depth sensor, for five depth values within the optimizated range.}
    \label{fig:psfs}
\end{figure}

In this section, we leverage the efficiency of the proposed neural optical model to enable the co-optimization of a millimeter scale metasurface with the parameters of a deep neural network for single-shot depth sensing. To the best of our knowledge, an optimization of this type has not been done previously with metasurface-based optics\footnote{We note that the co-design of a DOE with similar neural network architectures for depth sensing has been explored previously in \cite{wu2019phasecam3d,Ikoma:2021}}. We consider again a computational imaging architecture based on polarization-multiplexing as discussed in section \ref{sec:experiment} and schematically depicted in Figure \ref{fig:depth_pipeline}a. The working principle of this camera is based on depth-from-defocus (DfD). The metasurface forms two images, $\image_{0^\circ}$ and $\image_{90^\circ}$, at the photosensor, each with a distinct depth-dependent blur. The pair of images are then fed into a U-net which outputs a depth map of the scene. 

While it is possible for a sufficiently expressive U-net to generate depth predictions based on learned object statistics rather than cues from optical defocus, we encourage the latter by limiting the receptive field of the network. The receptive field size defines the number of pixels in the input image which are used to generate a depth prediction for each pixel at the output. The U-net architecture considered in this work is shown in Figure \ref{fig:depth_pipeline}b and utilizes 3x3 kernels throughout. The receptive field size is shown in the inset of Figure \ref{fig:depth-sensing}b. As the receptive field is small relative to the features in the image, we hypothesize that the U-net will be constrained to generate predictions based primarily on defocus.           
Notably, we emphasize that depth from defocus can be realized utilizing a single, 2D image as the input. Previous work on monocular depth estimation (MDE), however, have utilized neural network architectures with substantially more parameters (here, the network contains only 1.94 M parameters) and with a larger receptive field \cite{DBLP:journals/corr/abs-1812-11941, Liu2016LearningDF, NIPS2014_7bccfde7}. In consideration of a smaller computational architecture, we are motivated to utilize the capabilities of metasurfaces to encode two distinct PSFs simultaneously given recent demonstrations of a low computation, deterministic depth-from-differential-defocus algorithm based on two images \cite{Alexander_Focal_Flow, guo2019compact}. For constrained neural architectures, co-optimization based on two images should outperform single image estimates.    

Using the nanofin neural optical model and the Hankel-based radial Fresnel propagator, we inverse-design a 3 mm diameter, radially-symmetric metasurface in conjunction with the U-net parameters. The metasurface is initialized to focus light from depths of 0.3 m and 0.5 m at the photosensor for $0^\circ$ and $90^\circ$ linearly polarized light, respectively. For the optimization, we generate randomly synthesized scenes with the foreground and segmentation masks assembled from the Freiburg-Berkeley motion segmentation dataset \cite{OB14b} and the background from the COCO dataset \cite{DBLP:journals/corr/LinMBHPRDZ14}. Ground-truth depth maps consist of slanted planes. With both the scene and true depth as model inputs $\{(\scene_k, Z_{true, k})\}_{k=1,2,...}$, we utilize a training set of approximately 5000 scene-depth pairs and minimize the L1 loss via,    
\begin{equation}
    \arg\min_{\optparamSet, \compparamSet} \sum_k \left\| \text{U-Net} \left( \image_{0^\circ}(\scene_k), \image_{90^\circ}(\scene_k) \right) - Z_{true, k}\right\|,
\end{equation}
where $I_i$ is a polarization-dependent, rendered image similar to Equation \ref{eq:TPS_argmin} but utilizing the accelerated, slanted-plane depth rendering algorithm discussed in Section \ref{ssec:rendering_model}. Photon noise is added to the captured images \cite{Hasinoff2014PhotonPN}. 

In Figure~\ref{fig:depth-sensing}, we display the simulated performance of the optimized snapshot depth sensor for a test scene. The two predicted images produced by the metasurface (displayed in images a,b in the same figure) are in focus for different depths. The U-net successfully learns to recover a depth map of the scene based on the two image input, and in Figure \ref{fig:depth-sensing}f, we show a 2D histogram of the depth estimation performance across the test set containing 50 scenes. Across the optimized depth range, we find good performance. 

The initial and trained PSFs for the two polarization states are shown in Figure \ref{fig:psfs} (the PSF for one depth is extracted and magnified in Figure \ref{fig:depth-sensing}e). Remarkably, we observe that while the PSFs do change during training, the differences are relatively subtle. This suggests that the U-net architecture considered here, although relatively small compared to traditional MDE architectures, is still powerful enough to work in conjunction with the optics without having to depend heavily on finely-tuning the PSFs. Moreover, we believe that yet smaller and simpler neural architectures may be utilized in conjunction with two images and may be needed to realize the full potential of co-optimization rather than co-design for DfD with metasurfaces.

\section{Conclusion and Future Outlook}
\label{sec:conclusion}
In this work, we present a complete and auto-differentiable design framework for the co-optimization of metasurfaces and computational algorithms. We provide multiple, efficient and validated implementations for all stages of the design pipeline: field propagation, metasurface optical transformations, and rendering. While part of this work's contribution is the coherent synthesis and integration of prior research (some disjoint to the topic of metasurfaces), we also introduce the principle of the neural optical model--the usage of a multilayer perceptron to differentiably map nanoscale shapes to their local optical response. We discuss in detail the benefits of the neural optical model as an implicit representation, relative to auto-differentiable field solvers, and benchmark the performance against alternative approaches. We find that the neural optical model yields state of the art accuracy while being highly generalizable; consequently, it enables a new path forward for end-to-end metasurface design. 

In addition to the neural model, we also propose two new, theoretical metasurface-based computational imaging systems and demonstrate the usage of the framework to train them. We leverage the polarization multiplexing ability of a metasurface to capture two images in a single shot and from the same perspective, a feat which cannot be easily done with a single optic by other means. We then utilize this functionality to demonstrate a path for compact, incoherent opto-electronic image processing based on two-pupil synthesis and for efficient depth sensing based on a small receptive field U-net and depth from differential defocus. 

The source code for the framework is released as an open source package to the community, in addition to pre-trained, ready-to-use neural models. We believe that there is substantial room for further explorations and demonstrations including extending the MLP approach to higher dimensional cell shapes. New and specialized, adaptive sampling algorithms will likely be needed in order to efficiently generate the required training data for the MLP, as pre-computing the optical response for all parameter combinations will be infeasible for complicated cells. Moreover, the simple cell libraries utilized in this work enabled a straightforward method to impose constraints on the metasurface parameters. The development or implementation of alternative techniques is needed when the allowed values for the MLP inputs are conditional on one another. Previous research on dispersion engineering suggests that these higher dimensional cells are required in order to more freely engineer the optical response with respect to incident wavelength \cite{broadband_weiting, doi:10.1126/sciadv.abe4458}.     

While co-optimization of optical hardware and computational parameters is not a new idea, it is our opinion that the development of co-designed metasurface based systems remains in a nascent stage. The potential in applying end-to-end design of multiple images alongside modern and emerging techniques in image processing may likely continue to lead to the discovery of vision systems with substantially reduced computational costs.

\section{Acknowledgments}
\label{sec:ccknowledgements}
The authors thank Zhaoyi Li for review and helpful comments on this manuscript. This work was supported by the NSF IIS Award 1900847.

\appendix

\section{Neural Optical Model Additional Information}
\label{sec:supp_neural_optical}
In this section, we provide additional figures for the evaluation and analysis of the neural optical model. Discussion is provided in the main text. In Figure \ref{fig:supp_FinTrans_collage} and \ref{fig:supp_FinPhase_collage}, the transmission and phase predictions for x- and y-polarized light incident on a 350 nm cell with a centered nanofin structure is displayed for several trained neural optical models along with the alternative ERBF and multivariate polynomial models. In Figure \ref{fig:supp_nanocylinderMLP}, predictions on the nanocylinder cells are shown against the FDTD computed optical response. The mean absolute error, FLOPs, and number of parameters of all models tested in this work (some not visually shown in the figures) are listed in Table \ref{tab:supp_all_optical_models}.    

\begin{table}
    \centering
    \caption{Performance Per Cell Evaluation for All Surrogate Models}
    \label{tab:supp_all_optical_models}
    \small
    
    \begin{tabular}{|
    >{\columncolor[HTML]{EFEFEF}}l |
    >{\columncolor[HTML]{FFFFFF}}c |
    >{\columncolor[HTML]{FFFFFF}}c |
    >{\columncolor[HTML]{FFFFFF}}c |
    >{\columncolor[HTML]{FFFFFF}}c |}
    \hline
    
    Optical Model & \cellcolor[HTML]{EFEFEF}$\#$ Parameters & \cellcolor[HTML]{EFEFEF}FLOPs & \cellcolor[HTML]{EFEFEF} MAE Test Set$^3$ \\\hline
  
    Poly-14-cylinder &  360 & 717 k & 0.18\\ 
    Poly-17-cylinder &  513 & 1.02 k & 0.14\\ 
    Poly-23-cylinder &  900 & 2.3k & 0.095\\ 
    Poly-30-cylinder &  1.40 k & 2.8k & 0.082\\ \hline
    
    ERBF-32-cylinder &  0.2 k & 0.5 k & 0.097\\ 
    ERBF-64-cylinder &  0.4 k & 1 k & 0.047\\ 
    ERBF-128-cylinder &  0.9 k & 2 k & 0.029\\ 
    ERBF-256-cylinder &  1.8 k & 4 k & 0.026\\
    ERBF-512-cylinder &  3.6 k & 8 k & 0.029\\ \hline
    
    NO-D32-cylinder & 1 k & 2.8 k & 0.074 \\ 
    NO-D64-cylinder & 5 k & 10 k & 0.043 \\ 
    NO-D128-cylinder & 17 k & 36 k & 0.024 \\ 
    NO-D256-cylinder & 68 k & 138 k & 0.018 \\ \hline

    Poly-5-fins & 336 & 666 & 0.20 \\
    Poly-8-fins &  990 & 1.97 k & 0.13 \\ 
    Poly-11-fins &  2.18 k & 4.36 k & 0.09\\ 
    Poly-15-fins &  4.90 k & 9.8 k & 0.069\\ \hline

    ERBF-128-fins &  1.5 k & 3 k & 0.062 \\ 
    ERBF-256-fins &  3 k & 7 k & 0.040 \\ 
    ERBF-512-fins &  6 k & 14.0k & 0.036 \\ 
    ERBF-1024-fins &  12 k & 28 k & 0.032 \\
    ERBF-2048-fins &  25 k & 55 k & 0.026\\ \hline
    
    NO-D32-fins & 1 k & 3 k & 0.068\\ 
    NO-D64-fins & 5 k & 10 k & 0.047\\ 
    NO-D128-fins & 18 k& 37 k & 0.035\\ 
    NO-D256-fins & 68 k & 139 k & 0.025 \\ 
    NO-D512-fins & 267 k & 540 k & 0.021 \\
    NO-D1024-fins & 1.05 m & 2.13 m & 0.019 \\ \hline
    
    RCWA-($512^2,49$) & NA & 363 m & 0.062 \\ 
    RCWA-($512^2,81$) & NA & 1.62 b & 0.055 \\ 
    RCWA-($512^2,121$) & NA & 5.38 b & 0.051 \\ \hline

    \end{tabular}
\end{table}

\begin{figure*}
    \centering
    \includegraphics[width=1.0\textwidth]{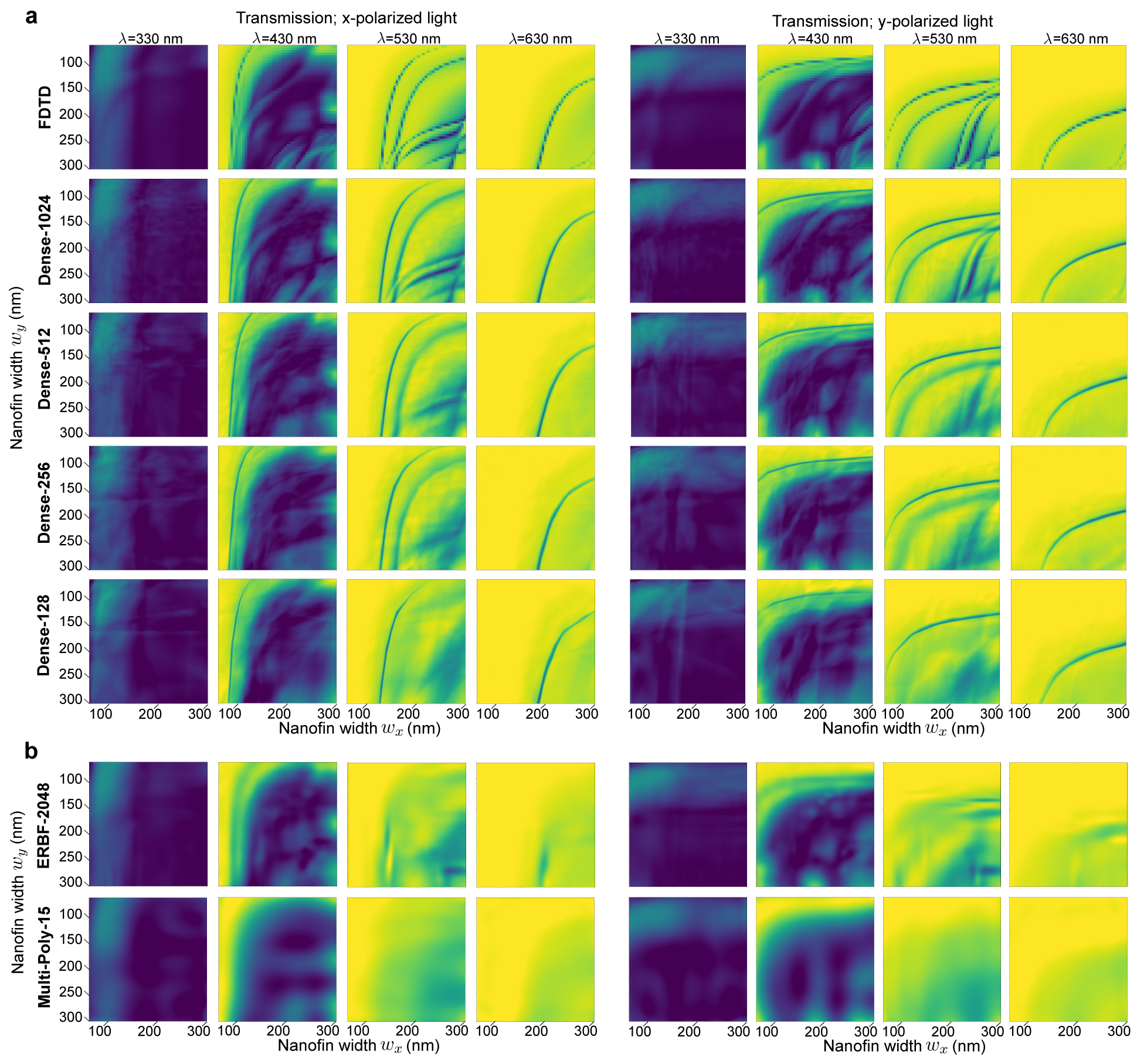}
    \caption{Similar to Figure \ref{fig:Nanofin_mlp}a in the main paper; Neural optical models with different numbers of parameters in each hidden layer are trained on nanofin cells. The predicted transmittance for different cells is displayed (upsampled at 4x the resolution of the training dataset) and contrasted against the ground-truth FDTD results. Each model (each row) has two hidden, dense layers. The number of neurons in each layer corresponds to the number in the label on the left column, i.e. Dense-1024 has two hidden, dense layers of 1024 neurons.}
    \label{fig:supp_FinTrans_collage}
\end{figure*}

\begin{figure*}
    \centering
    \includegraphics[width=1.0\textwidth]{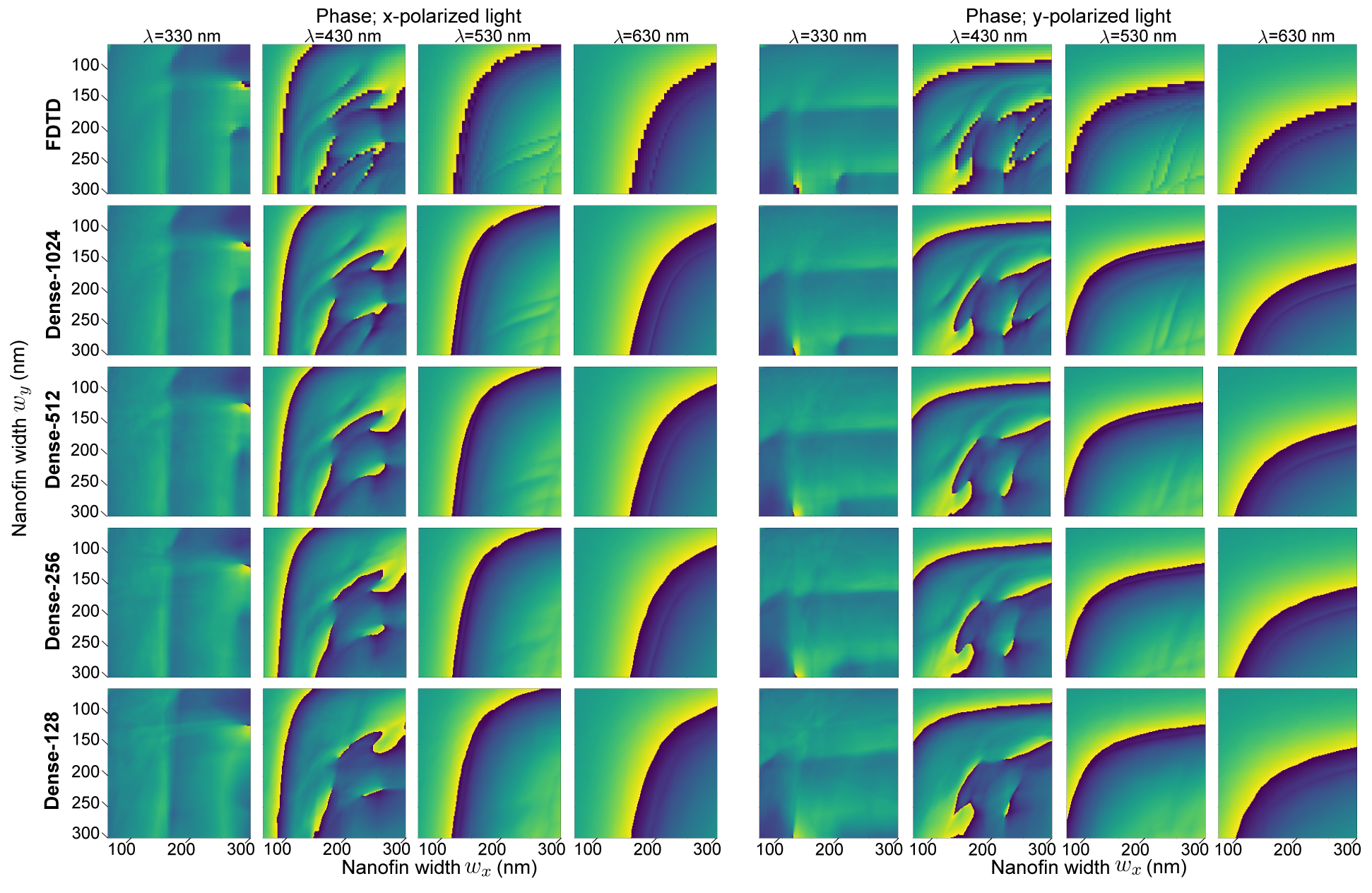}
    \caption{Similar to Figure \ref{fig:Nanofin_mlp}a in the main paper and complementary to Figure \ref{fig:supp_FinTrans_collage}. See the caption there for more details. The MLP predicted phase is contrasted against the FDTD ground-truth data. The performance of the MLP with different numbers of neurons in each hidden layer is shown.}
    \label{fig:supp_FinPhase_collage}
\end{figure*}

\begin{figure*}
    \centering
    \includegraphics[width=1.0\textwidth]{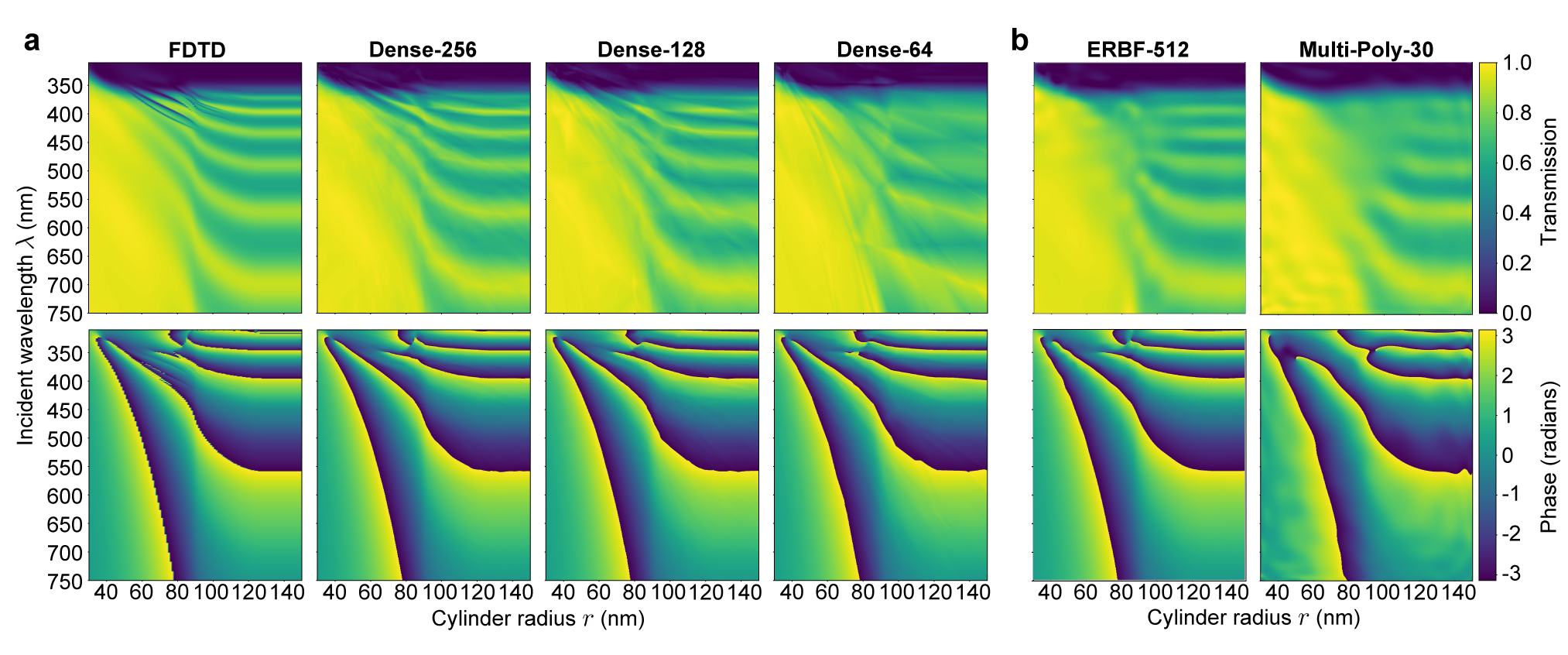}
    \caption{MLP-predictions for the transmittance (top row) and phase (bottom row) imparted by a 180 nm cell with 600 nm tall nanocylinders of different radii placed at the center (as depicted in Fig 1b of the main document). The FDTD optical response is shown along with different, pre-trained neural optical models labeled as Dense-N; the models consists of two hidden, dense layers with N neurons in each. As with the nanofins, we find a continuous trade-off between model accuracy and number of parameters (see Table 1 in the main document for quantitative measures).}
    \label{fig:supp_nanocylinderMLP}
\end{figure*}

\section{Propagated Field Validation}
\label{sec:supp_propagation_validation}
\begin{figure*}
    \centering
    \includegraphics[width=1.0\textwidth]{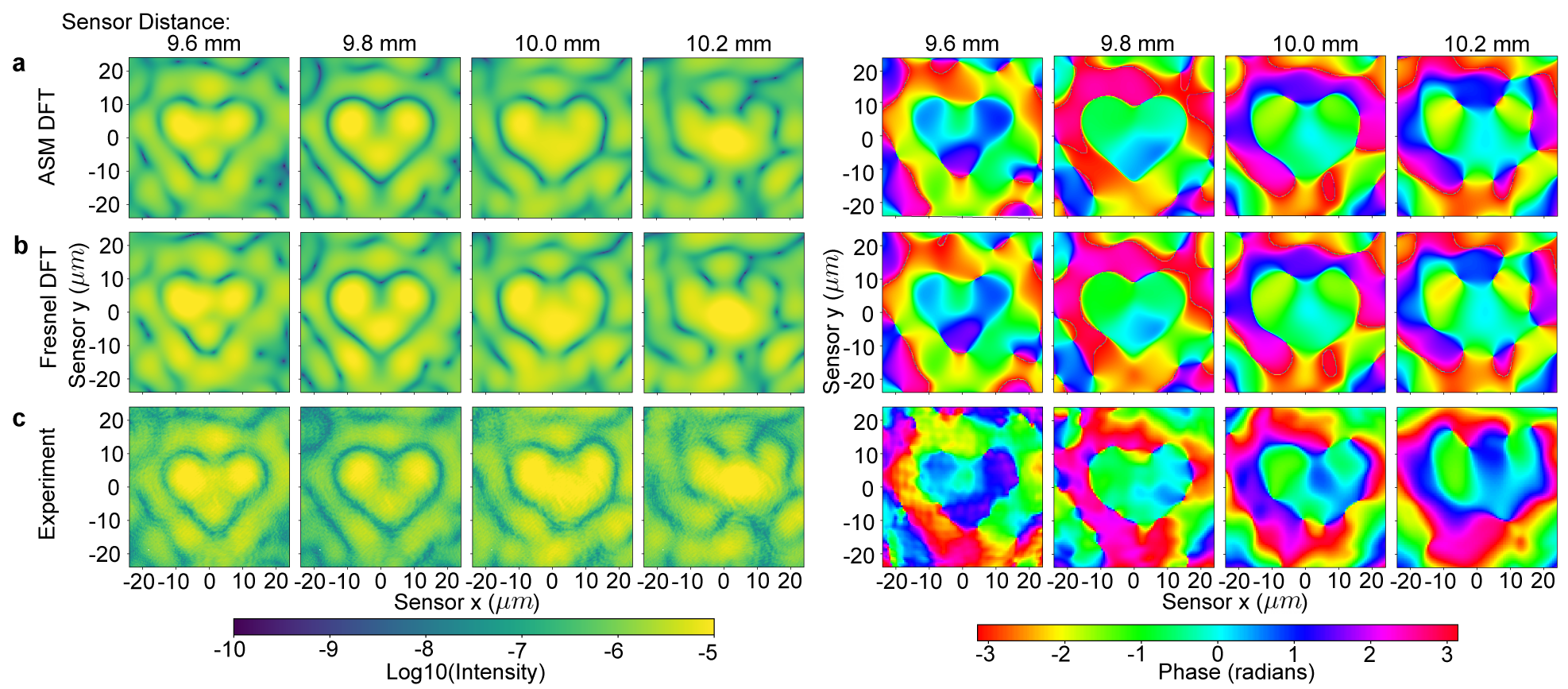}
    \caption{Complementary to Figure \ref{fig:HeartSingularity} in the main document. The predicted intensity and phase of the field at the sensor plane is computed with \sellnameShort~using (a) the angular spectrum method and (b) the Fresnel diffraction method. In (c), the experimental measurements taken by Lim et al. and published in \cite{SWDLim_Singularity_NatCommun_2021} are shown. Four different metasurface to sensor distances are considered.}
    \label{fig:Supp_heart_hologram}
\end{figure*}

As introduced in Section \ref{ssec:prop_val} of the main document, we present one validation for the field propagators included in \sellnameShort~by analyzing the intensity produced at the focal plane of various matelenses (metasurfaces which are designed to focus light as diffraction-limited lenses). We consider a field just after the metalens, assuming plane wave incidence, according to the hyperbolic phase profile introduced in Equation \ref{eq:focusing_phase} of the main text:
\begin{equation}
        \wave(\lr, 0^+) = \exp\left(\frac{-\imag 2\pi}{\wl}\left(f-\sqrt{\lr^2 + f^2}\right)\right)
\end{equation}
where f is the focal length and $\lr$ denotes radial coordinates on the metasurface plane, $\lr = \sqrt{\lx^2 + \ly^2}$. In this analysis, we assume that the metalens implements the required phase-delay at each cell exactly and without modulating the field intensity; In other words, we consider only the propagation of this ideal field to the sensor plane. In the main text of Section \ref{ssec:prop_val}, this assumption is eased when the metasurface is inverse designed to discover the placement of nanofin cells that focus light. 

Here, we consider the field on a grid after a 100 $\mu m$ diameter metalens with square cells of 350 nm. The grid at the sensor plane is finely sampled at 50 nm. We consider instantiations of metalenses that are designed to focus light at sensor distances of 100, 150, and 200 $\mu$m and for incident wavelengths $\wl$ of 380, 532, and 700 nm, respectively. The intensity at the sensor plane for each case is then computed and a central slice through the intensity profiles are shown in Figure \ref{fig:supp_metalens_focusing}. These sensor distances are chosen as they correspond to different values of the imaging numerical aperture, $\text{NA} = r/f$, where r is the radius of a circular aperture on the metalens. The NA is a useful metric as it relates to the range of angles over which the system operates. Moreover, it is understood that the Fresnel approximation (one assumption of paraxial optics) breaks down for large NA values. In this regime, we expect only the angular spectrum method (ASM) to produce the correct, propagated fields. 

The calculations in Figure \ref{fig:supp_metalens_focusing} are plotted alongside the theoretic Airy disk profile. We find good agreement between the theoretic Airy intensity and the propagated fields. For NA larger than 0.30, the Fresnel approximation becomes less accurate while the radial and 2D ASM calculations are still in strong agreement. 

\begin{figure*}
    \centering
    \includegraphics[width=0.9\textwidth]{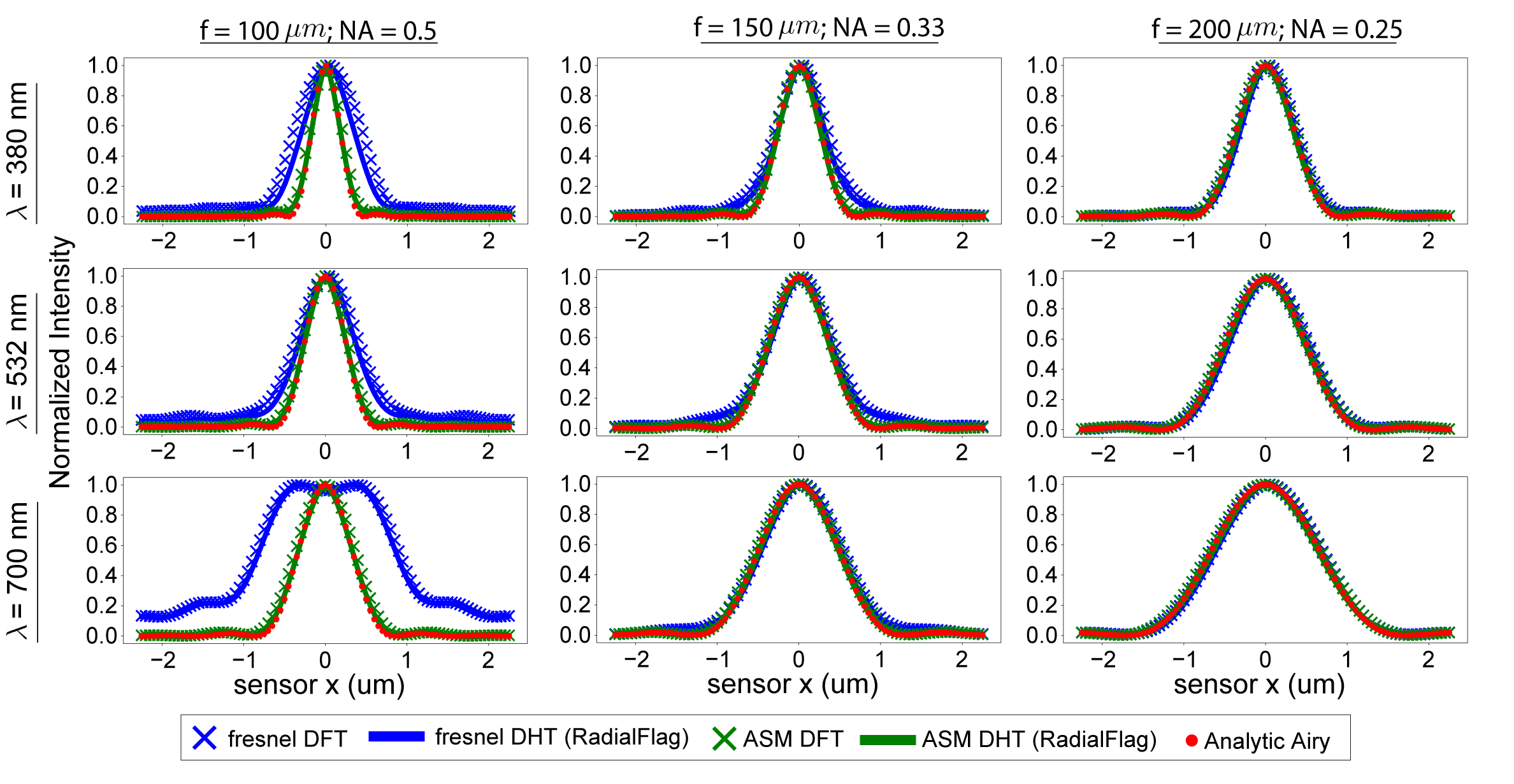}\\
    \caption{See the text in Appendix \ref{sec:supp_propagation_validation} for more details. Each graph in the figure corresponds to a different metalens instantiation, which imparts the hyperbolic phase profile required to focus an incident plane wave of a particular wavelength $\wl$ to a particular sensor distance. A radial slice of the intensity profile at the sensor plane is shown, as computed by the four propagators included in \sellnameShort. The analytic intensity profile prescribed by the Airy disk is also plotted in red. The simulated calculations should match closely to the Airy profile with disagreement only for the Fresnel method at large NA.}
    \label{fig:supp_metalens_focusing}
\end{figure*}

\section{Incoherent Image Differentiation Additional Details}
\label{sec:supp_tps_details}
To better understand the spatial frequency filter imposed by the metasurface based system, it is useful to consider the optical transfer function (OTF). By taking the Fourier transform of both sides of Equation \ref{eq:TPS} in the main text, we obtain the frequency-space relation:
\begin{equation}
    \widetilde{\image}_{0^\circ} - \alpha \widetilde{\image}_{90^\circ} = \widetilde{\texture}\cdot \ft{\psf_{0^\circ}-\alpha \psf_{90^\circ}}= \widetilde{\texture}\cdot\ft{\psf'},
\end{equation}
where $\widetilde{\image}$ and $\widetilde{\texture}$ denotes the Fourier transform of $\image$ and $\texture$ respectively. The significance of this representation is that it reveals that the frequency content in the captured images $\image$ corresponds to the frequency content in the pinhole image $\texture$ modulated by a spatial frequency filter. This spatial frequency filter is the Fourier transform of the PSF and is referred to as the optical transfer function (OTF) by convention. Just as we assign an effective, net PSF $\psf'$ to the opto-electronic system, we may define an effective, net OTF via:
\begin{equation}
    \text{Net OTF} = \ft{\psf'}.
\end{equation}
The operation of image differentiation is well defined in frequency space as it corresponds to a particular, complex-valued spatial frequency filter. We may then compare the net OTF of the optimized system to this ideal filter. 

The net PSF and the net OTF for the optimized imaging system that applies a first derivative to the pinhole image (section \ref{ssec:image_differentiation}) is displayed in Figure \ref{fig:sup_tps_Firstderiv}. As noted in the caption, the OTF is consistent to the definition of a first derivative for low spatial frequencies. Disagreement will always occur at high frequencies and this can be understood by recognizing that any physical, optical system must have a high spatial frequency cut-off. This cutoff may be set by the pixel size or by the diffraction-limit and sufficiently high frequency components in a scene cannot be imaged. In other words, the OTF must go to zero for high frequencies. The net PSF and the net OTF for the second-derivative imaging system is similarly displayed in Figure \ref{fig:sup_tps_2nd_deriv_zoom}.

\begin{figure*}
    \centering
    \includegraphics[width=1.0\textwidth]{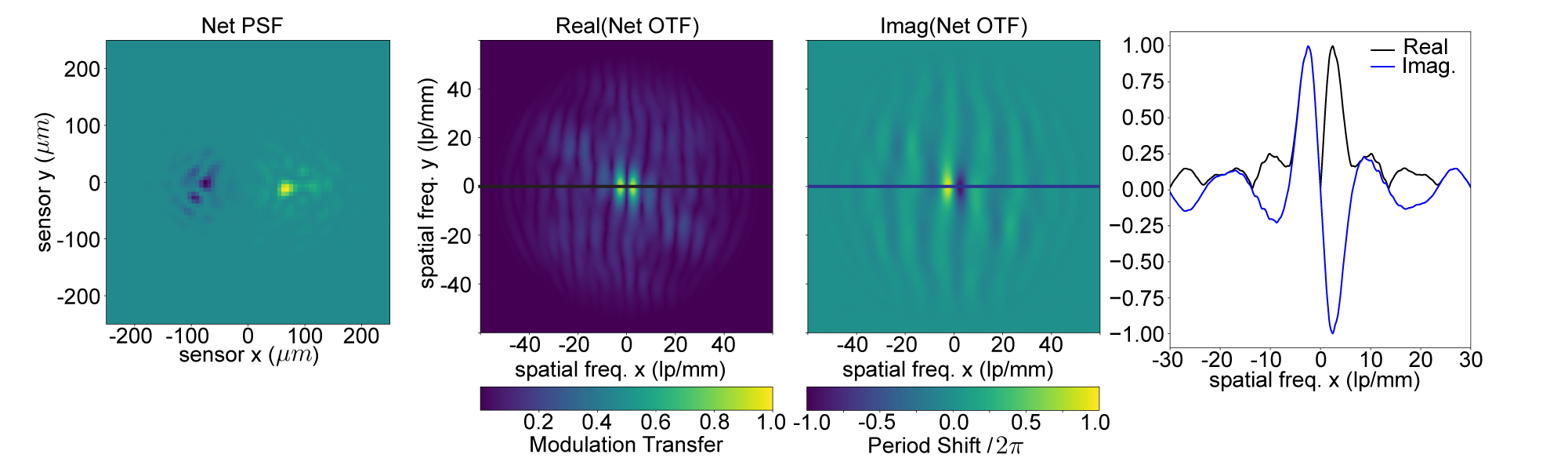}
    \caption{The net PSF $\psf'$ for the optimized imaging system introduced in section \ref{ssec:image_differentiation} (Figure \ref{fig:tps_firstDeriv}) of the main text is reproduced. This effective PSF results in the first-derivative image of a scene and corresponds to an object to metasurface distance of 0.75 m. To the right, we show the real and imaginary parts of the OTF obtained by taking the Fourier transform of $\psf'$. Notably, there is zero modulation transfer for the zero-spatial frequency component. For a small range around $f_x=0$, the real part is approximately linear while the imaginary part is anti-symmetric, consistent to the definition of the first derivative.}
    \label{fig:sup_tps_Firstderiv}
\end{figure*}

\begin{figure*}
    \centering
    \includegraphics[width=1.0\textwidth]{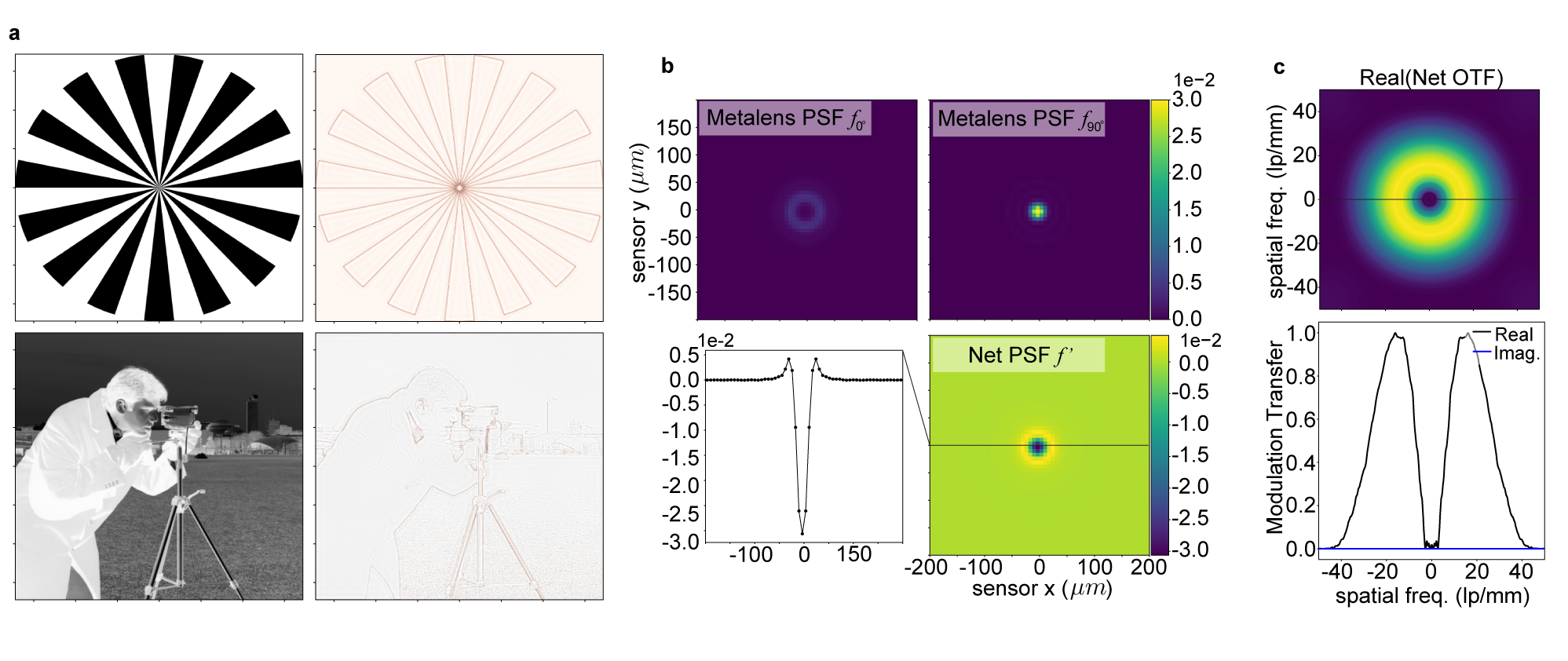}
    \caption{(a) Enlarged images of the pinhole image and the net image from Figure \ref{fig:tps_2ndDeriv} in the main document. The metasurface visual system enables a second-derivative rendering of the scene for two FLOPs per pixel. (b) The two PSFs for the optimized metasurface are shown alongside the effective, net PSF $\psf'$. The discovered effective PSF has a similar shape to the conventional Laplacian of Gaussian kernel used for edge-detection. (c) The real part of the net OTF is shown. The imaginary part of the OTF is approximately zero for all spatial frequencies. The real part of the net OTF has an approximately quadratic shape near $f_x=0$. These features are consistent to the frequency definition of a second derivative.}
    \label{fig:sup_tps_2nd_deriv_zoom}
\end{figure*}

\bibliographystyle{ACM-Reference-Format}
\bibliography{doc_bibfile}

\end{document}
\endinput
https://www.overleaf.com/project/62b089f18622d74751305502https://www.overleaf.com/project/62b089f18622d74751305502https://www.overleaf.com/project/62b089f18622d74751305502